\documentclass[sigconf]{acmart}
\AtBeginDocument{%
  \providecommand\BibTeX{{%
    Bib\TeX}}}

\copyrightyear{2025}
\acmYear{2025}
\setcopyright{acmlicensed}\acmConference[CCS '25]{Proceedings of the 2025 ACM SIGSAC Conference on Computer and Communications Security}{October 13--17, 2025}{Taipei, Taiwan}
\acmBooktitle{Proceedings of the 2025 ACM SIGSAC Conference on Computer and Communications Security (CCS '25), October 13--17, 2025, Taipei, Taiwan}
\acmDOI{10.1145/3719027.3765138}
\acmISBN{979-8-4007-1525-9/2025/10}

\usepackage{algorithmic}
\usepackage{graphicx}
\usepackage{textcomp}
\usepackage{xcolor}
\usepackage{comment}
\usepackage{listings}
\usepackage{wrapfig}
\usepackage{proof}
\usepackage{multicol}
\usepackage{comment}
\usepackage{ulem}
\usepackage{pifont}
\usepackage{bbm}
\usepackage{tikz-cd}
\newtheorem{definition}{Definition}
\newtheorem{theorem}{Theorem}
\newtheorem{lemma}{Lemma}
\usepackage{amsmath}
\usepackage{balance} 

\lstdefinestyle{Imp}{
  language= {},
  aboveskip=3mm,
  belowskip=3mm,
  showstringspaces=false,
  columns=flexible,
  basicstyle={\small\ttfamily},
  numberstyle=\tiny\color{gray},
  keywordstyle=\bfseries,
  stringstyle=\color{mauve},
  breaklines=true,
  breakatwhitespace=true,
  escapeinside={(*}{*)},
  morekeywords={EVENT,WHERE,THEN,END, AWAIT, IF, FI, ATOM, record,axiomatization,where,and},
  morecomment=[f][\color{gray}][0]{//},
  tabsize=3
}

\lstdefinestyle{CStyle}{
    language=C,
    basicstyle=\ttfamily\footnotesize,
    keywordstyle=\color{blue}\bfseries,
    commentstyle=\color{cyan}\itshape,
    stringstyle=\color{red},
    numberstyle=\tiny\color{gray},
    stepnumber=1,
    numbersep=10pt,
    backgroundcolor=\color{white},
    tabsize=4,
    showstringspaces=false,
    captionpos=b,
    breaklines=true,
    breakatwhitespace=false,
    escapeinside={(*}{*)},
}

\newcommand{\cmdfont}[1]{\textbf{#1}} 

\newcommand{\defi}{\triangleq}

\newcommand{\cmdbasic}[1]{\cmdfont{Basic} \ {#1}}
\newcommand{\cmdseq}[2]{\cmdfont{Seq} \ {#1} \ {#2}}
\newcommand{\cmdcond}[3]{\cmdfont{Cond} \ {#1} \ {#2} \ {#3}}
\newcommand{\cmdwhile}[2]{\cmdfont{While} \ {#1} \ {#2}}
\newcommand{\cmdawait}[2]{\cmdfont{Await} \ {#1} \ {#2}}
\newcommand{\cmdnondt}[1]{\cmdfont{Nondt} \ {#1}}
\newcommand{\cmdnone}{\bot}

\newcommand{\progpar}[2]{#1 \parallel \cdots \parallel #2}

\newcommand{\sys}[2]{\cmdfont{let } #1 \cmdfont{ in } #2}
\newcommand{\sysdefault}{\sys{\Pi}{\progpar{\symbprog}{\symbprog}}}

\newcommand{\symbprog}{P}

\newcommand{\symbevt}{e}

\newcommand{\symbState}{S}

\newcommand{\symbstate}{s}

\newcommand{\symbevtctx}{x}
\newcommand{\symbConf}{\Delta}
\newcommand{\symbconf}{\delta}

\newcommand{\symbcore}{\kappa}

\newcommand{\symbDomain}{\mathcal{D}}
\newcommand{\symbdomain}{d}

\newcommand{\symbSM}{\mathcal{M}}
\newcommand{\symbAction}{A}
\newcommand{\symbaction}{a}
\newcommand{\symbactions}{as}

\newcommand{\symbdomf}{dome}
\newcommand{\symbpolicy}{\mathtt{Info}}
\newcommand{\symblabel}{l}
\newcommand{\symbguard}{g}
\newcommand{\symbSys}{W}
\newcommand{\symbpart}{Part}

\newcommand{\tran}[1]{\stackrel{{#1}}{\longrightarrow}}

\newcommand{\dsim}[1]{\stackrel{{#1}}{\sim}}

\newcommand{\interf}{\leadsto}

\newcommand{\ssequidom}[3]{{#1}\stackrel{#2}{\bumpeq}{#3}}

\newcommand{\simuli}{\preceq}
\newcommand{\refine}{\sqsubseteq}

\newcommand{\lstate}{\symbstate}
\newcommand{\hstate}{\sigma}
\newcommand{\lRely}{R}
\newcommand{\lGuar}{G}
\newcommand{\hRely}{\mathbb{R}}
\newcommand{\hGuar}{\mathbb{G}}

\newcommand{\stutter}{\tau}

\makeatletter
\newcommand{\superimpose}[2]{%
  {\ooalign{$#1\@firstoftwo#2$\cr\hfil$#1\@secondoftwo#2$\hfil\cr}}}
\makeatother

\newcommand{\ninterf}{\mathrel{\mkern-3mu\not\mkern3mu\leadsto}}

\def\BibTeX{{\rm B\kern-.05em{\sc i\kern-.025em b}\kern-.08em
    T\kern-.1667em\lower.7ex\hbox{E}\kern-.125emX}}

\begin{document}

\title{Generalized Security-Preserving Refinement for Concurrent Systems}

\author{Huan Sun}
\affiliation{%
  \institution{Zhejiang University}
  \city{Hangzhou}
  \country{China}
}
\email{huansun@zju.edu.cn}

\author{David Sanán}
\affiliation{%
  \institution{Singapore Institute of Technology}
  \country{Singapore}
}
\email{david.sanan@singaporetech.edu.sg}

\author{Jingyi Wang} 
\authornote{Corresponding authors: Jingyi Wang and Wenhai Wang}
\affiliation{%
  \institution{Zhejiang University}
  \city{Hangzhou}
  \country{China}
}
\email{wangjyee@zju.edu.cn}

\author{Yongwang Zhao} 
\affiliation{%
  \institution{Zhejiang University}
  \city{Hangzhou}
  \country{China}
}
\email{zhaoyw@zju.edu.cn}

\author{Jun Sun}
\affiliation{%
  \institution{Singapore Management University}
  \country{Singapore}
}
\email{junsun@smu.edu.sg}

\author{Wenhai Wang} 
\authornotemark[1]
\affiliation{%
  \institution{Zhejiang University}
  \city{Hangzhou}
  \country{China}
}
\email{zdzzlab@zju.edu.cn}

\renewcommand{\shortauthors}{Huan Sun et al.}

\begin{abstract}
Ensuring compliance with Information Flow Security (IFS)
is known to be challenging, especially for concurrent systems with large codebases such as multicore operating system (OS) kernels. Refinement, which verifies that an implementation preserves certain properties of a more abstract specification, is promising for tackling such challenges. However, in terms of refinement-based verification of security properties, existing techniques are still restricted to sequential systems or lack the expressiveness needed to capture complex security policies for concurrent systems.

In this work, we present a generalized security-preserving refinement technique, particularly for verifying the IFS of concurrent systems governed by potentially complex security policies. 
We formalize the IFS properties for concurrent systems and present a refinement-based compositional approach to prove that the generalized security properties (e.g., intransitive noninterference) are preserved between implementation and abstraction. The key intuition enabling such reasoning, compared to previous refinement work, is to establish a step-mapping relation between the implementation and the abstraction, which is sufficient to ensure that every paired step (in the abstraction and the implementation, respectively) is either permitted or prohibited by the security policy. We apply our approach to verify two non-trivial case studies against a collection of security policies.
Our proofs are fully mechanized in Isabelle/HOL, during which we identified that two covert channels previously reported in the ARINC 653 single-core standard also exist in the ARINC 653 multicore standard. We subsequently proved the correctness of the revised mechanism, showcasing the effectiveness of our approach. 
\end{abstract}




\maketitle

\section{Introduction}
Ensuring the security and integrity of information is both highly desirable and challenging for modern software systems, particularly for complex concurrent systems like multicore operating system (OS) kernels. These systems often present a broad attack surface, making them susceptible to exploitation by malicious actors seeking unauthorized access to sensitive information. A formally verified security proof can provide strong guarantees that such exploits are impossible. 
However, the current state-of-the-art remains inadequate in achieving such a high level of assurance, which often involves significant costs or is simply infeasible.

Refinement \cite{refine} is a powerful technique in formal verification that ensures a more concrete or lower-level system preserves the essential properties of a more abstract or higher-level specification. Refinement has been successfully applied to verify real-world large-scale codebases, including OS kernels \cite{sel41,sel42,certikos,ccal,ucos}, file systems \cite{filesystem1,filesysystem2,filesystem3}, and cloud hypervisors \cite{sekvm,sekvm1,kvm_secure}. However, in terms of refinement-based verification of security properties, existing techniques are still restricted to sequential systems or lack the expressiveness needed to capture complex security policies for concurrent systems. Indeed, applying refinement to verify general security properties in practical concurrent systems such as multicore OS kernels faces several fundamental technical challenges:

\begin{enumerate}
    \item \textbf{Preserving Security Properties}: It is well-known that refinement does not inherently preserve security \cite{notpreserve1,notpreserve2,mcertikos}. For concurrent systems, most refinement techniques focus on preserving safety \cite{rgsim,csim,Igloo} or liveness properties \cite{ccal,lili1,trillium}, but not security properties. Ensuring that refinement preserves security properties requires additional considerations, particularly in handling information flow and interference across concurrent executions.
    \item \textbf{Concurrency:} For sequential systems, security-preserving refinement has been extensively studied \cite{sel4_noninterference,sel4_noninterference2,mcertikos,serval,nickel,arinc653_ifs,zhao_refinement}. However, these techniques often fail to generalize to the concurrent setting. As noted by \cite{kvm_secure}, intermediate executions in concurrent systems that are visible to other parallel components may be hidden by refinement, which 
    can lead to unintended information leakage.
    \item\textbf{Support General Security Policies}: 
    Recently, transparent trace refinement \cite{kvm_secure} has been proposed to track updates to shared data in concurrent systems, ensuring that intermediate updates that remain observable across parallel components are not concealed by refinement. However, this approach is restricted to isolation-based policies, where strict noninterference is enforced. In contrast, many practical concurrent systems require controlled information release (e.g., downgrading or declassification mechanisms) to accommodate real-world security requirements. Achieving security-preserving refinement for general security policies remains an open challenge.   
\end{enumerate}

These challenges highlight the need for a security-preserving refinement framework capable of handling general security policies for concurrent systems. In this work, we aim to bridge this significant gap from the following key dimensions.

\textbf{\textit{Generalized Security Policy.}} 
Note that a standard way to prove IFS is to formulate the security property as noninterference \cite{goguen82}, which ensures that actions at low-security levels (public) remain independent of those at higher levels (secret), preventing adversaries from inferring secrets or influencing decisions through unauthorized means. 
While effective, this strict prohibition of information flow from high to low levels can be impractical for systems that require controlled declassification of information. To support controlled declassification, language-based security \cite{declassification,veronica,verdeca} guarantees that attackers can only learn secrets through explicitly specified declassification commands. More generally, intransitive noninterference \cite{rushby81,rushby92,otherifs} 
allows information to flow from low to high, from high to a declassifier, and then from the declassifier to low (while still prohibiting direct high-to-low flow). In this work, we primarily focus on intransitive noninterference, as it 
avoids the need to rewrite a system in a specific language, and 
provides a general and clear security specification, making it more suitable for system verification\footnote{It is possible to apply our refinement technique to language-based security though. We will provide a more detailed discussion in Section \ref{sect:relate}.}, as demonstrated in prior work on seL4 \cite{sel4_noninterference,sel4_noninterference2}, CertiKOS \cite{mcertikos,serval}, NiKOS \cite{nickel}, and ARINC 653 \cite{arinc653_ifs,zhao_refinement}.  

\textbf{\textit{Security-preserving Refinement for Concurrency.}} The concept of unwinding conditions, originally introduced by Rushby \cite{rushby81, rushby92}, provides a compositional method for proving noninterference: instead of reasoning about the entire system trace, it suffices to check that each execution step maintains state indistinguishability unless explicitly permitted by the security policy. Many efforts have been made to preserve unwinding conditions \cite{sel4_noninterference,sel4_noninterference2,mcertikos,nickel}, but mostly only work for sequential systems. Recent work \cite{sekvm,sekvm1,kvm_secure} has introduced transparent trace refinement that successfully propagates unwinding conditions for concurrent systems restricted to isolation policies. 
Notably, if the abstract specification violates state indistinguishability (even when permitted by the security policy), these approaches will fail to provide guarantees. 

In this work, we present a new unwinding-preserving simulation between implementation and abstract specification to reason about intransitive noninterference for concurrent systems. The key idea is to establish a step-mapping relation between the implementation and the abstract specification. \textit{A step of execution in the implementation is either a silent step that will preserve state indistinguishably relation or has a corresponding step in the abstract specification that is equivalent concerning the security policy, i.e., a step in implementation breaks the state indistinguishably relation only when the corresponding step in the abstract specification does, and both actions are either allowed or disallowed by the security policy}. Consequently, if the abstract specification is proved secure, so is the implementation.

We apply our approach to verify two non-trivial case studies: the practical inter-partition communication (IPC) mechanism of ARINC 653 multicore \cite{ARINC653p1_4}, a leading safety standard for partitioning operating systems in multicore environments (security is not verified before), and a benchmark sealed-bid auction system requiring declassification to show its effectiveness. We identified that two covert channels previously reported in the ARINC 653 single-core standard \cite{arinc653_ifs} also exist in the ARINC 653 multicore standard. We subsequently proved the security of the revised version and formalized all our results in Isabelle/HOL\footnote{The source files of Isabelle/HOL are available at \url{https://github.com/IS2Lab/Refine_IFS}
}. Notably, we found in our case studies that supporting security-preserving refinement does not significantly increase the proof burden compared to standard refinement techniques. This makes our approach promising to flexibly extend existing proofs to support security reasoning of new systems.

To summarize, we make the following main contributions:
\begin{enumerate}
    \item A novel simulation technique that preserves unwinding conditions for concurrent systems under general security policies, i.e., intransitive noninterference \cite{rushby81,rushby92} (supporting modeling of intentional data release and downgrading).
    To support compositional reasoning, we adopt a rely-guarantee simulation methodology, which decomposes the simulation proof of the entire system into proofs over individual parallel components.
    \item An end-to-end security proof, fully formalized in the Isabelle proof assistant, for two nontrivial concurrent systems with distinct security policies: the IPC mechanism of the ARINC 653 multicore standard and a sealed-bid auction system. During verification, we identified that two security violations previously reported in the single-core standard also exist in the multi-core version and formally proved the security of the revised version.
\end{enumerate}

\section{Overview}
\label{sect:ex}
We start by introducing an illustrative example showing potential security violations in the IPC mechanism of a multicore OS. We then examine the cause of failures for existing refinement techniques and present the key idea of our approach. 

\begin{figure}
    \centering
    \begin{multicols}{2}
    \begin{lstlisting}[style=CStyle]
struct msgq {
  size, max :: nat;
  que :: msg list;
  mutex l;}
struct kernel_state {
  cid :: tid;
  t_msg :: tid (* $\rightarrow$*) msgq
  t_cnt :: tid (* $\rightarrow$*) nat}
// recv function
msg recv(){
  msgq q = t_msg(cid);
  msg m = Null;
  lock(q->l)
  if (q->size > 0)
    msg m = q->que[size-1];
    q->size --;
    unlock(q->l);
    return m;
}

//Thread t1 on CPU 1
void T1(msg m) {send(t2, m)}
//Thread t2 on CPU 2
void T2(){
  msg m:= recv();
  send(m, t3)}
//Thread t3 on CPU 
msg T3() {
  return recv()}
    \end{lstlisting}   
    \end{multicols}
    \begin{multicols}{2}
    \begin{lstlisting}[style = CStyle]
// insecure send
int send(tid id, msg m){
  t_cnt(id) ++;  
  if check(cid, id, policy)
    msgq q = t_msg(id);
    lock(q->l)
    if (q->size = q->max)
      return -1
    else
      q->que[size] = m;
      q->size ++;
    unlock(q->l)
  else
    t_cnt(id) --;
  return 0;}
// secure send
void send(tid id, msg m){
  if check(cid, id, policy)
    msgq q = t_msg(id);
    lock(q->l)
    if (q->size < q->max)
      q->que[size] = m;
      q->size ++;
    unlock(q->l)}
    \end{lstlisting}
    \end{multicols}
    \caption{IPC Mechanism in multicore OS}
    \label{fig:example}
\end{figure}

\subsection{An Illustrative Example}
Consider a multicore OS kernel supporting parallel execution of system calls. We want to prove that one thread can communicate with another through communication channels under a specified security policy.

\textbf{\textit{Example. }}
Fig.~\ref{fig:example} illustrates the IPC mechanism adapted from a commercialized multicore OS--ARINC 653 multicore. A message queue is a data structure containing a list of messages $que$ where $size$ represents the number of messages in the queue and $max$ is the queue's capacity. Since the message queue is shared between different threads, a mutex $l$ is used to manage access, ensuring that only one thread can access the queue at a time. The kernel state comprises three elements: $cid$ represents the thread ID of the currently executing thread, $t\_msg\_queue$ is a mapping from thread IDs to message queues, representing the message buffer of each thread, $t\_cnt$  is a counter that tracks the total number of send attempts made to each thread $id$. The kernel provides two system calls: $\text{send}(id, m)$ tries to send the message $m$ to the thread $id$. Before enqueuing the message into the buffer of thread $id$, the system call checks the security policy to ensure that the information transfer from the calling thread to $id$ is permitted. $\text{recv}()$ retrieves a message from the calling thread’s message buffer. 

There are two differences between the insecure and secure implementations of the send function. (1) The insecure implementation increments the counter $t\_cnt$ before checking the security policy. If the policy disallows the send operation, the counter is then decremented to revert the change. (2) The insecure implementation informs the sender when the message queue is full (i.e., $size = max$), while the secure implementation simply discards the message.

\subsection{Security Formulation}
The security within the system is defined through security policy $\interf$, which specifies the permitted information flows among different threads. Below is an example security policy,  where the only permitted information flow is $t_{1} \interf t_{2}$, $t_{2} \interf t_{3}$, and $t_{3} \interf t_{1}$. 

\begin{center}
\begin{tabular}{cccc}
 & $t_{1}$ & $t_{2}$ & $t_{3}$ \\
 $t_{1}$ & -    &   \checkmark   &  \ding{55}  \\
 $t_{2}$ &  \ding{55}    & -    &   \checkmark   \\
 $t_{3}$ &   \checkmark   &   \ding{55}   & -   
 \end{tabular}    
\end{center}

This policy is intransitive, i.e., $t_{1} \interf t_{2}$ and $t_{2} \interf t_{3}$ does not imply $t_{1} \interf t_{3}$. This policy can be used to specify the requirement that sensitive information must pass through the downgrading domains before being released. For example, the security policy in Fig. \ref{fig:example} specifies all message transferred from $t_{1}$ to $t_{3}$ must be mediated by $t_{2}$, direct communication from $t_{1}$ to $t_{3}$ is explicitly prohibited.

To specify what a thread can observe and how information may flow through different threads, we use the state indistinguishability relation $\dsim{t}$ to specify for each thread $t$, the set of states that appear to be indistinguishable. For the example in Fig. \ref{fig:example}, if the secure send is implemented, then a thread $t$ can only observe its message queue $t\_msg(t)$, that means all states with the same $t\_msg(t)$ can not be distinguished by the thread $t$, thus the state indistinguishability relation is defined as:

$$ \symbstate \dsim{t} \symbstate' \defi \symbstate(t\_msg(t)) = \symbstate'(t\_msg(t))$$

If the insecure send is implemented, then a thread $t$ can additionally observe the counter $t\_cnt(t)$ and \textit{full status} of another thread, then the state indistinguishability relation is defined as:

\begin{equation*}
\begin{aligned}
     \symbstate \dsim{t} \symbstate' \defi \ & \symbstate(t\_msg(t)) = \symbstate'(t\_msg(t)) \wedge \symbstate(t\_cnt(t)) = \symbstate'(t\_cnt(t)) \wedge\\
    & \forall t'. \ t \interf t' \implies  \symbstate (full(t')) = \symbstate'(full(t'))
\end{aligned}
\end{equation*}

Intuitively, if $t_{i} \ninterf t_{j}$, every step performed by the thread $t_{i}$ can not be detected by the thread $t_{j}$, i.e., preserves the state indistinguishability. Now consider thread $t_{1}$ trying to send messages to $t_{2}$, and they are running on $CPU_1$ and $CPU_2$, respectively. As the security policy shows, we want to ensure only $t_1$ can send information to $t_2$, i.e., $ t_1 \interf t_2$ and $t_2 \ninterf t_1$. 

Two covert storage channels can arise if the insecure implementation is used. (1) Prematurely updating the shared counter $t\_cnt$ introduces a covert channel. Although the final state may be rolled back, the intermediate update to $t\_cnt$ reveals that a message send attempt was made—even if it was not permitted by the security policy. In a concurrent setting, this intermediate update can be observed by thread $t_1$, leading to unintended information leakage. (2) The \textit{full status} of the message queue can be exploited as a covert channel that leaks information from $t_2$ to $t_1$. Since $t_1$ can observe the full status of the message queue, while $t_2$ can change this status, information can be implicitly transferred to $t_2$, violating the security policy.

In our setting, thread scheduling is purely nondeterministic. For a given concurrent system $t_1 \parallel \cdots \parallel t_n$, the scheduler nondeterministically chooses one enabled thread $t_i$ to execute a single step, and the state is updated accordingly. A thread is considered enabled if its next command is not a lock acquisition, or it is a \texttt{lock}$(l)$ command and the lock $l$ is currently available. There are no fairness or priority constraints, and the choice of thread is not influenced by any secret or observable information. Similarly, if multiple threads simultaneously attempt to acquire the same lock, the winner is selected nondeterministically.

As a result, there is no internal timing channel \cite{internal,commcsl} in our model. This design choice simplifies the security reasoning: since scheduling behavior is independent of secrets, we do not need to explicitly consider timing behaviors in our verification. Our work focuses instead on covert storage channels, where security is defined in terms of state-based observability, and timing behavior is not considered part of the observable state.



\subsection{Security-Preserving Refinement}
Refinement~\cite{refine} is a powerful verification technique that involves developing an abstract specification of a system (simple and easy to reason about) and then proving that a concrete implementation refines this specification. Fig.~\ref{fig:abstraction} shows the abstraction for the \text{send} and \text{recv} functions, which atomically ($\langle \symbprog \rangle$ means $\symbprog$ is executed atomically) enqueue a message into the receiver thread's message queue and dequeue a message from the calling thread's message queue, respectively. The key goal of refinement is to ensure that certain properties (e.g., safety, liveness, or security) proven for the abstract specification are preserved by the concrete implementation.

The standard refinement technique typically relies on a simulation relation $\alpha$ between the states of the concrete implementation and those of the abstract specification. The simulation requires that for every step in the implementation, there must exist zero or more steps in the abstraction such that the resulting states satisfy the simulation relation.  A simple simulation relation $\alpha$ might say that the message queues in the implementation and abstraction contain the same messages, ignoring other variables such as locks or counters. As long as each step in the implementation maintains this relation with the abstract state, we say that the implementation simulates the specification. For example, if the abstract specification ensures that the number of messages in the queue never exceeds the capacity, and if the simulation relation holds between the implementation and the abstraction, then this safety property is also satisfied by the implementation.


\textbf{\textit{Standard Refinement May Fail to Preserve Security.}} 
There has been significant effort \cite{rgsim,csim,Igloo,ccal,lili1,lili2,termpreserve1,termpreserve2,trillium,simuliris} to establish refinement for concurrent systems. However, these approaches are limited to proving certain safety or liveness properties, overlooking security properties. 

\begin{figure}
    \centering
    \includegraphics[width=0.9\linewidth]{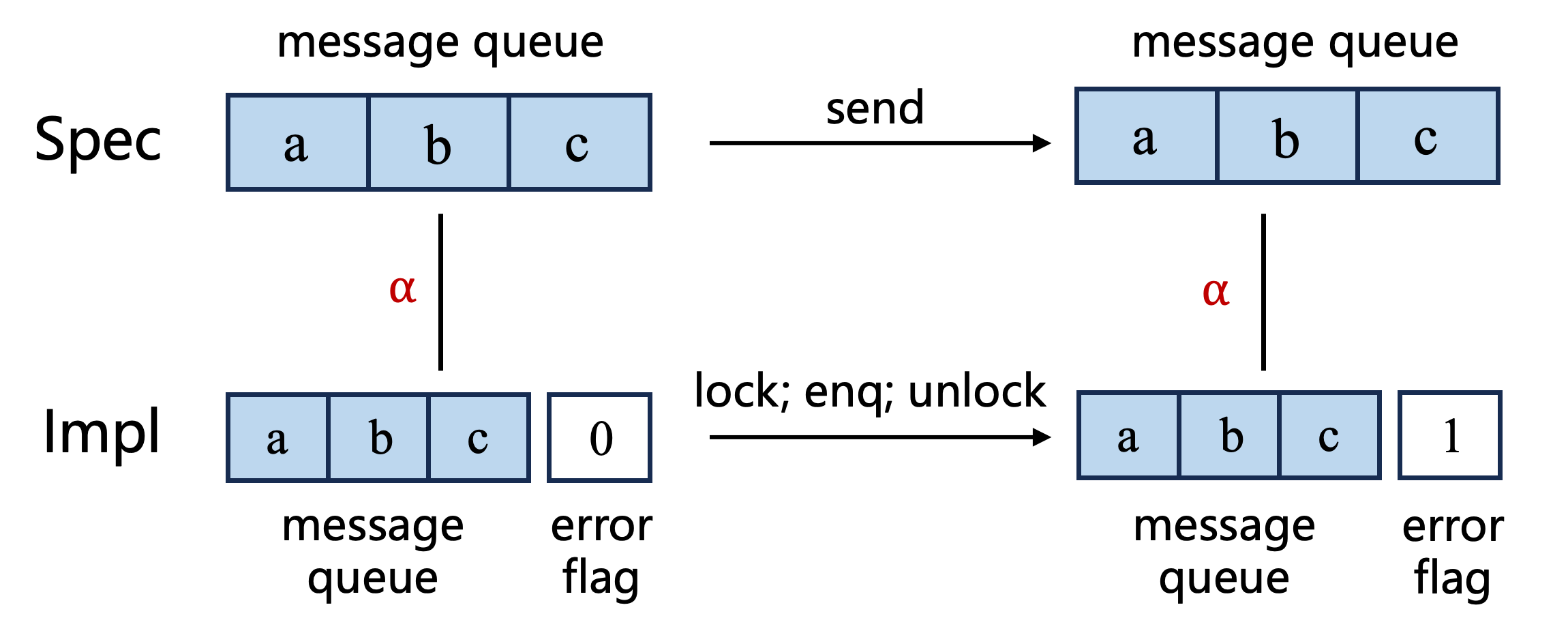}
    \caption{Security-Violating Refinement}
    \label{fig:recv}
\end{figure}

One issue, as noted in \cite{certikos}, is that simulation relations used in refinement often lack a direct connection to the security policy. Fig.~\ref{fig:recv} illustrates this problem. Suppose that the simulation relation $\alpha$ relates two states in which the message queues contain the same messages, but their error flags may differ. Using this simulation relation, it is easy to show that both secure and insecure $\mathsf{send}$ implementations refine the abstract specification in Fig. \ref{fig:abstraction}. However, this example fails to preserve security. The abstraction satisfies the security policy, as the sender can never infer the full status of the receiver, while the insecure implementation allows the sender to obtain information from the receiver, causing information leakage.

\begin{figure}
    \centering
    \begin{multicols}{2}
    \begin{lstlisting}[style=CStyle]
// abstraction for send 
void send(tid id, msg m){
  (*$\langle$*) if check(cid, id, policy)
    msgq q = t_msg(id);
    if (q->size < q->max)
      q->que[size] = m;
      q->size ++ (*$\rangle$*)}

// abstraction for recv
msg recv(){
  (*$\langle$*)msgq q = t_msg(cid);
  msg m = Null;
  if (q->size > 0)
    msg m = q->que[size-1];
    q->size --;
    return m(*$\rangle$*)}
    \end{lstlisting}
    \end{multicols}
    \caption{Abstraction for Send and Read Function}
    \label{fig:abstraction}
\end{figure}

\begin{figure*}
  \centering
  \begin{minipage}{0.29\textwidth}
    \centering
    \includegraphics[width=\textwidth]{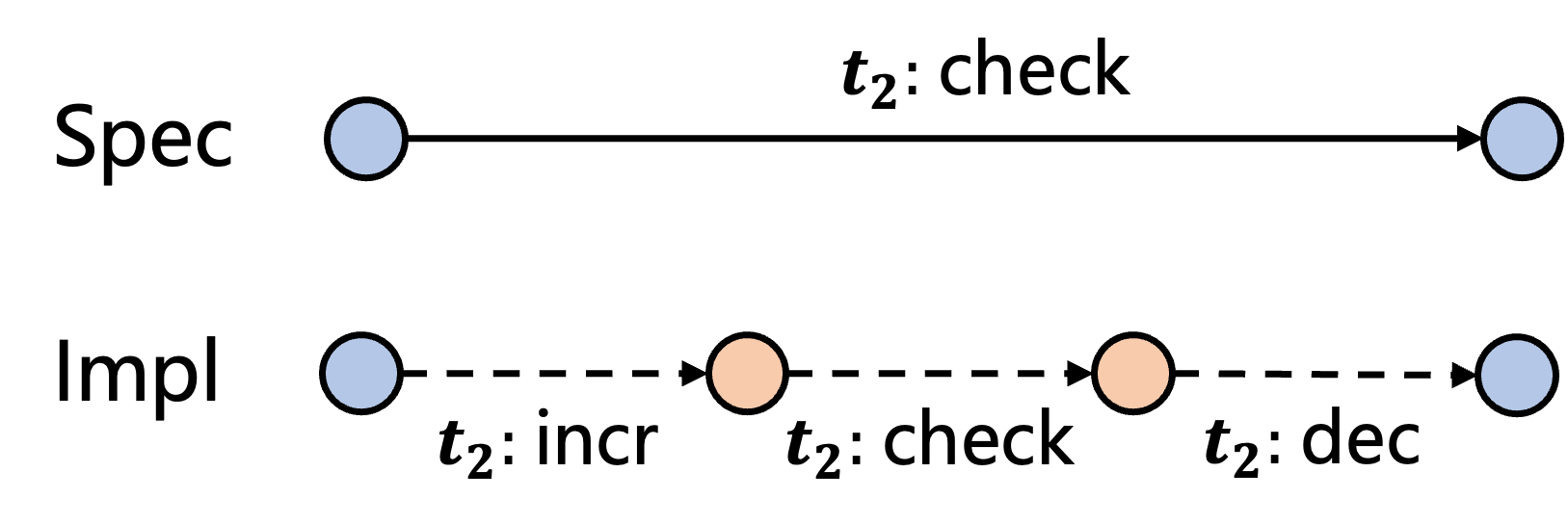}
    \caption{Sequential Refinement}
    \label{fig:seq}
    \end{minipage}%
  \begin{minipage}{0.3\textwidth}
    \centering
    \includegraphics[width=\textwidth]{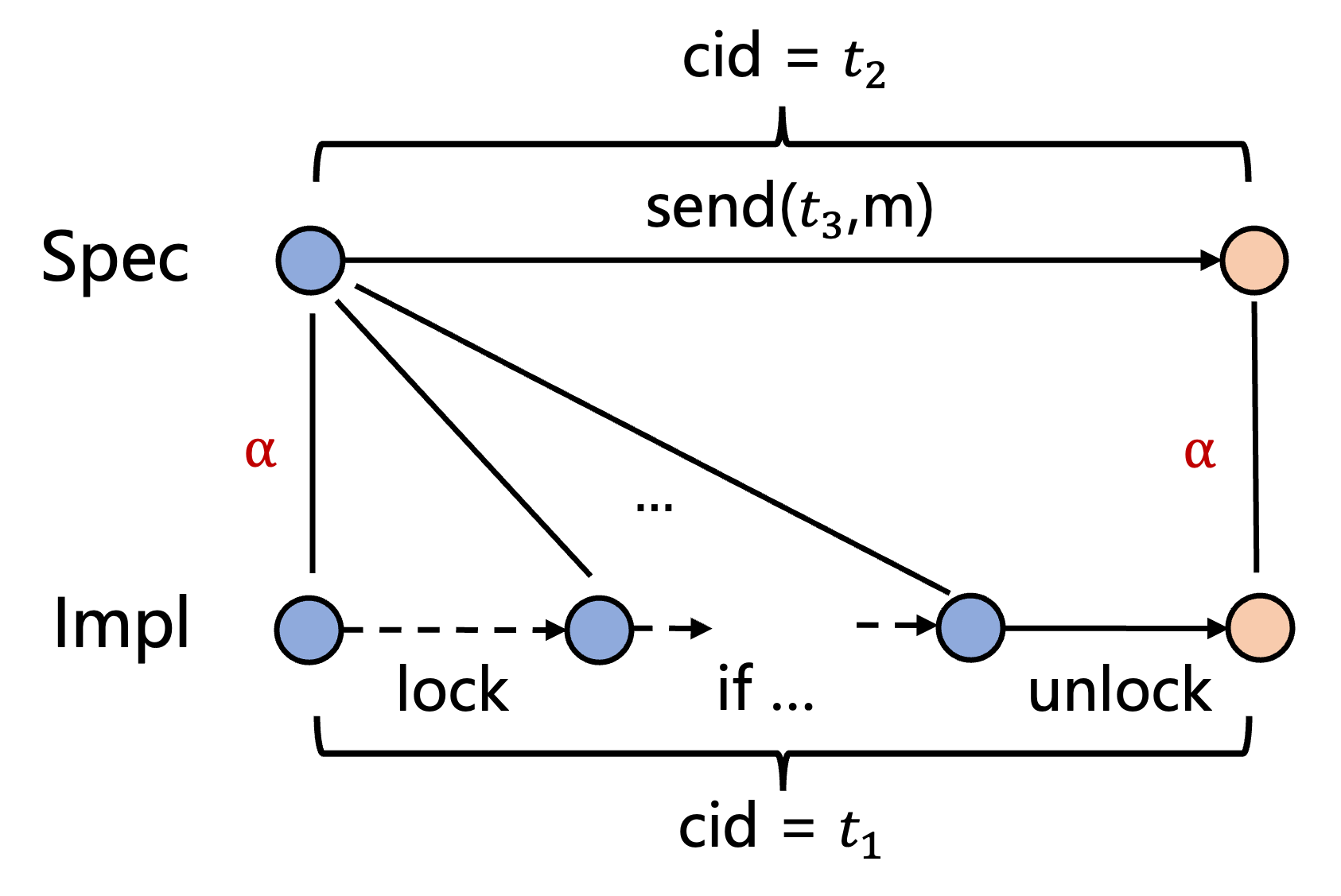}
    \caption{Transparent Trace Refinement}
    \label{fig:send}
  \end{minipage}%
  \hspace{1em}
  \begin{minipage}{0.3\textwidth}
    \centering
    \includegraphics[width=\textwidth]{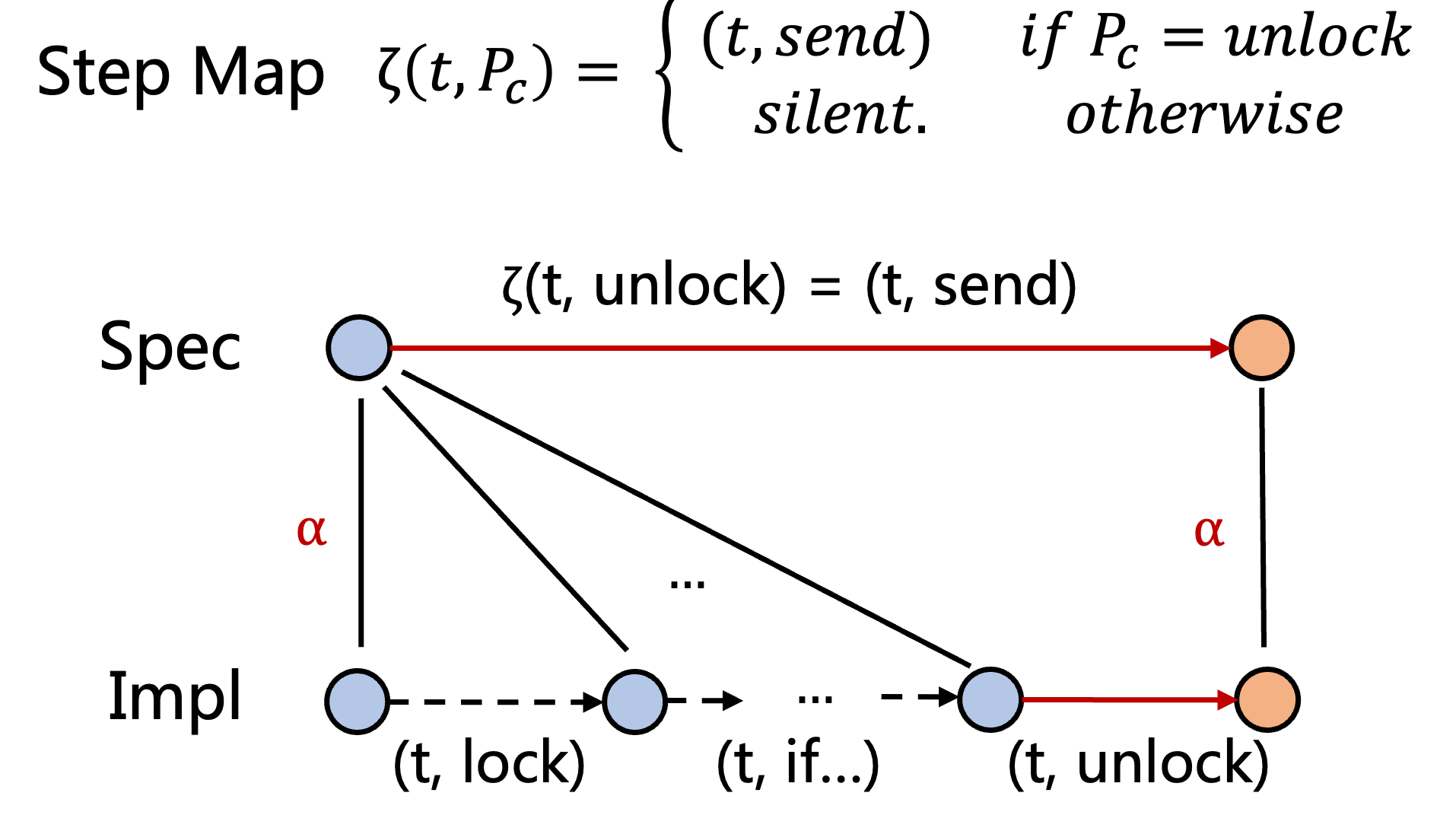}
    \caption{Our approach}
    \label{fig:frame}
  \end{minipage}
\end{figure*}

\textbf{\textit{Concurrency Challenge.}}
Security-preserving refinement has been extensively studied in sequential systems \cite{sel4_noninterference,sel4_noninterference2,mcertikos,serval,nickel,arinc653_ifs,zhao_refinement}. However, these techniques fail to preserve security in the presence of concurrency. As highlighted by \cite{kvm_secure}, the core issue is that intermediate executions (though potentially visible to other parallel components) can be hidden by refinement. For instance, the sequential security-preserving refinement technique cannot detect the counter channel illustrated in Fig.\ref{fig:example}. As shown in Fig. \ref{fig:seq}, although the complete execution of the insecure $\text{send}(t_1, m)$ function invoked by $t_{2}$ does not violate security policy, its intermediate update to the counter $t\_cnt$ is visible by $t_{1}$, leading to information leakage. Since the sequential refinement technique abstracts away these intermediate steps, it fails to capture this security breach, thereby failing to preserve the intended security guarantees.

\textbf{\textit{Fail to Support General Security Policies.}} Transparent trace refinement \cite{sekvm, sekvm1, kvm_secure} ensures that whenever an atomic step in the implementation disrupts state indistinguishability, there exists a step in the abstraction that does the same. Consequently, if the abstraction is verified to always preserve state indistinguishability, the implementation does as well.
Transparent trace refinement is sufficient to preserve security under an isolation policy, where every atomic step must preserve state indistinguishability. However, for systems that permit information transfer (e.g., intransitive security policy), it may fail to ensure security as well. Figure~\ref{fig:send} illustrates a scenario where transparent trace refinement accepts the simulation relation but compromises security. Suppose the simulation relation $\alpha$ relates two states in which the message queues contain the same message, but their current thread ID $cid$ may differ. In the implementation, thread $t_{1}$ enqueues a message to $t_{3}$'s message queue due to the wrong implementation of the $\mathsf{check}$ function. It is possible to find a related state in the abstraction where $cid = t_{2}$ and also enqueues a message to $t_{3}$'s message queue. Although both actions have the same effect---violating the state indistinguishability relation of $t_{3}$---the former is disallowed, while the latter is permitted.

The fundamental issue here is that \textbf{\textit{transparent trace refinement guarantees only that any violation of state indistinguishability in the implementation also appears in the abstraction. However, it does not establish a precise correspondence between the specific actions responsible for these violations.}} In other words, transparent trace refinement fails to ensure the responsible steps in the implementation and abstraction are both allowed or disallowed by the policy. As a consequence, it may accept insecure implementations that violate the intended security guarantees.

To support intentional data release, \cite{sekvm,sekvm1,kvm_secure} introduces a data oracle to non-declaratively encode the security policy. Instead of exposing actual states, the data oracle modifies the semantic model by replacing declassified sensitive data with non-sensitive data, allowing declassification to be proved in terms of an isolation policy. For example, when information is intentionally released from $t_1$ to $t_2$, the data oracle pretends that the message queue of $t_{2}$ appears identical, regardless of the message sent by $t_1$, effectively obscuring $t_1$'s influence on the state observed by $t_2$. However, we argue that this approach complicates the specification's implications, as the additional preconditions or axioms introduced by data oracles must be carefully audited, and it remains unclear what security guarantees hold for the actual system. The reason stems from the transparent trace refinement only guarantees that security properties are preserved when the abstract specification preserves state indistinguishability, which is too restrictive for systems that allow certain information exchange. 

\textbf{\textit{Our Approach.}} Figure \ref{fig:frame} illustrates the core idea of our security-preserving refinement framework for concurrent systems under general security policies. To avoid intermediate execution effects being concealed by refinement, a step in our system is modeled as an atomic computation that occurs on a single parallel component. The goal is to ensure that every step in the implementation either maintains state indistinguishability or has a corresponding step in the abstract specification that has the same observational effect. This is achieved by two key components: state invariant relation $\alpha$ \footnote{We use $\alpha$ as it is essentially a type of simulation.}, and a step mapping function $\zeta$, to establish the correspondence between the states and steps of the implementation and the abstract specification. 

We require the state invariant relation $\alpha$ to preserve state indistinguishability to resolve the incompatibility between the simulation relation and state indistinguishability, i.e., If $(\lstate_{1}, \hstate_{1}) \in \alpha$ and $(\lstate_{2}, \hstate_{2}) \in \alpha$, then $\lstate_{1} \ \stackrel{d}{\sim_{a}} \ \lstate_{2} \equiv \hstate_{1} \ \stackrel{d}{\sim_{c}} \ \hstate_{2}$ for all $\symbdomain$.
This ensures our refinement approach can distinguish between secure and insecure implementations in Fig. \ref{fig:recv}. If the state invariant relation $\alpha$ preserves state indistinguishability, it must relate two states in which the message queues contain the same messages and the error flags are consistent. An insecure implementation breaks the simulation relation when the error flag becomes inconsistent, thereby failing to refine the abstract specification.

The step mapping function $\zeta$ classifies each implementation step into one of two categories: a \emph{silent step}, which must preserve state indistinguishability, i.e., it never breaks state indistinguishability, or an \emph{action step}, which is mapped to a specific corresponding step in the abstraction. This requirement ensures that intermediate actions observable by other components are not abstracted away. For instance, the refinement shown in Fig.~\ref{fig:seq} is invalid because the \texttt{incr} step, which lacks a corresponding abstract step and should therefore be silent, leaks information and thus breaks the state indistinguishability relation.

An action step is mapped to a specific abstract step, both must adhere to the same security constraints. In the example of Fig. \ref{fig:example}, we use a tuple $(t, P_{c})$ to represent a program step $P_{c}$ performed by thread $t$. We use the step mapping function $\zeta$ to indicate the mapping between steps in abstraction and implementation. All steps before the unlock operation are silent steps. These steps can be shown to preserve the state indistinguishability relation because, while a thread holds the lock, other threads cannot observe the state of the message queue it is accessing. The \texttt{unlock} step, however, is mapped by $\zeta$ to an abstract step that atomically enqueues a message into the destination thread’s message queue. Although such steps may break state indistinguishability, they can be shown to be either both allowed or disallowed by the security policy. 
In the example of Fig. \ref{fig:send}, if $t_{1}$ truly sends the message to $t_{3}$ in the implementation, then there exists a send action performed by $t_{1}$ in the abstract specification that does the same thing. Consequently, it will be rejected by the security policy, showing that security is preserved through refinement. 

In practice, the step mapping function is often straightforward to define. For linearizable concurrent programs \cite{linear} (e.g., lock-based programs), all steps except the linearization point should be unobservable by other threads. Thus, the step mapping function typically maps the linearization point (e.g., the unlock operation in lock-based implementations) to the corresponding atomic step in the abstract specification. Furthermore, as discussed in Section~\ref{sect:exp}, our method introduces minimal additional proof burden compared to standard refinement approaches.

\section{Basic Technical Setting}
\label{sect:ba}
This section provides the system model and security policy for concurrent systems. Then we define security-preserving refinement between the implementation and abstract specification, showing that the security properties are preserved through refinement. 

\paragraph{System Model} 

We model the concurrent system as a state machine where each action models some atomic computation taking place on a single parallel component; concurrency is realized by the nondeterministic interleaving of steps across all parallel components:

\begin{definition}[State Machine]
The system model of concurrent systems $\symbSM=\langle \symbState, \symbAction, step, \symbstate_0 \rangle$ consists of the following components:

\begin{enumerate}
    \item a set of states $\symbState$
    \item a set of actions $\symbAction$
    \item a step function $step: \symbState \times \symbAction \rightarrow \mathcal{P}(S)$
    \item a initial state $\symbstate_{0}$
\end{enumerate}
\end{definition}

An action $a\in A$ transitions the system from one state to another. In the context of a multicore OS, an action can be either an atomic step $P_{c}$ of a system call invoked by a thread $t$, represented as $(t, P_{c})$, or a schedule action, represented as $sched$. 

\paragraph{Security Policy}

The information flow within the system is specified by the information flow configuration $\symbpolicy$:
\begin{definition}[information flow configuration]
An information flow configuration $\symbpolicy = \langle \symbDomain, \interf, dom, \sim \rangle$ of the system $\symbSM$ consists of the following components:
\begin{enumerate}
    \item a set of domains $\symbDomain$
    \item a security policy $\interf \ \subseteq \symbDomain \times \symbDomain$
    \item a domain function $dom$
    \item a state indistinguishability relation $\sim \ \subseteq \symbDomain \times \symbState \times \symbState$
\end{enumerate}
    
\end{definition}
A domain represents the authority of an action. In the context of multicore OS, the domain is represented as $\{t_{1}, \cdots, t_{k}\} \cup \{ scheduler\}$, which means that a domain can be a thread or the scheduler.

The security policy $\interf$  constrains the flow of information among domains, $\symbdomain_1 \interf \symbdomain_2$ means that information is allowed to flow from $\symbdomain_1$ to $\symbdomain_2$, where $\symbdomain_1$ is often regarded as a low-security domain and $\symbdomain_2$ a high one. $\ninterf$ is the complement relation of $\interf$. We call the policy intransitive \cite{rushby92} if the relation $\interf$ can be intransitive, i.e., $u \interf v$ and $v \interf w$ \emph{does not necessarily} imply $u \interf w$. Note that intransitive policy is a generalized version of the transitive policy, which allows us to support a broad range of policies applied in practical systems. The security policy for IPC mechanism presented in Fig. \ref{fig:example} is intransitive, as $t_{1} \interf t_{2}$, and $t_{2} \interf t_{3}$, but $t_{1} \ninterf t_{3}$. Another example is downgrading operations, as practical systems often need to intentionally release data to users (e.g., through an encryption module). An intransitive policy can capture this by specifying that information can flow from the system to the encryption module and from the encryption module to users, but never directly from the system to users.

We use $dom(a)$ to denote the domain that performs $a$. For example, the domain of a step execution of a system call is the current running thread, and the domain of a schedule action is the scheduler. In our model, the function $dom$ is defined as follows:

\[
\text{dom}(a) =
\begin{cases}
t, & \text{if } a = (t, P_{c}) \\
scheduler, & \text{if } a = sched
\end{cases}
\]
The state indistinguishability relation $\sim$ is a reflexive, symmetric, and transitive relation that specifies the set of states that are indistinguishable by a given domain. we use $ \symbstate_{1} \dsim{\symbdomain} \symbstate_{2}$ to denote $(\symbdomain, \symbstate_{1}, \symbstate_{2}) \in \ \sim$. 

\paragraph{Security-Preserving Refinement}
The standard proof technique of intransitive noninterference is by proving a set of unwinding conditions \cite{rushby92,rushby81,sel4_noninterference,sel4_noninterference2,mcertikos,arinc653_ifs,nickel}
that examines individual actions on the system. 


Given a system $\symbSM = \langle \symbState, \symbAction, step, \symbstate_0 \rangle$ and its associated information flow policy $\symbpolicy = \langle \symbDomain, \interf, dom, \sim \rangle$, unwinding conditions are formally defined as follows:

\begin{definition}[Local Respect - LR]
\begin{equation*}
\begin{aligned}
LR(\symbSM, \symbpolicy) \defi &\ \forall \symbaction \ \symbdomain \ \symbstate \ \symbstate'. \   \symbstate' \in step(\symbstate, \symbaction) \\ 
& \implies dom(\symbaction) \ninterf \symbdomain \implies \symbstate \dsim{\symbdomain} \symbstate'
\end{aligned}
\end{equation*}
\end{definition}

Intuitively, the Local Respect (LR) condition requires that an action $\symbaction$ executed in a state $\symbstate$ may affect state indistinguishability only for those domains to which the executing domain is permitted to transmit information. For example, the scenario illustrated in Fig.~\ref{fig:example} violates the LR condition under the insecure implementation. Specifically, although $t_{2} \ninterf t_{1}$, an action performed by $t_{2}$ can alter the full status of its message queue, thereby compromising the state indistinguishability of $t_{1}$.

\begin{definition}[Step Consistent - SC]
\begin{equation*}
\begin{aligned}
& SC(\symbSM, \symbpolicy) \defi \forall \symbaction \ \symbdomain \ \symbstate_1 \ \symbstate_2 \ \symbstate_{1}' \ \symbstate_{2}'. \\
& \symbstate_1 \dsim{\symbdomain} \symbstate_2 \implies \symbstate_1' \in step(\symbstate_1, \symbaction)  \implies  \symbstate_2' \in step(\symbstate_2, \symbaction)  \\
&  \implies ((dom(\symbaction) \interf \symbdomain) \implies (\symbstate_1 \dsim{dom(\symbaction)} \symbstate_2)) \implies \symbstate_1' \dsim{\symbdomain} \symbstate_2' \\
\end{aligned}
\end{equation*}
\end{definition}

Intuitively, the SC condition requires that executing an action on two indistinguishable states preserves state indistinguishability for all domains if those states are indistinguishable from the perspective of the action’s domain. For example, consider two states $s_{1}$ and $s_{2}$ where $s_{1} \dsim{t_{3}} s_{2}$—indicating that the message queue of $t_{3}$ is identical in both states, SC condition requires the resulting states of performing the $\text{send}(t_3, m)$ action remain equivalent for $t_{3}$.

Intransitive noninterference can be derived from unwinding conditions. We omit the formal definition of intransitive noninterference due to page limit, and readers can refer to \cite{rushby81,rushby92} and the appendix for a detailed definition:
\begin{theorem}[Unwinding Theorem of Noninterference]
\label{thm:ut_noninfl}
\[ \text{SC}(\symbSM, \symbpolicy) \wedge \text{LR}(\symbSM, \symbpolicy) \implies \text{NI}(\symbSM, \symbpolicy) \]
\end{theorem}

Notice that many other IFS properties can be derived from unwinding conditions, e.g., TA-security \cite{otherifs}, which guarantees that information will not be leaked through the order of actions. 
Unwinding conditions can also be used as a security property in their own right. We omit the descriptions of other properties here and refer to \cite{otherifs} for a detailed review.

Now we define the security-preserving refinement, which says unwinding conditions are preserved when refining an abstract specification into a concrete implementation:
\begin{definition}[Security-Preserving Refinement]
We say that $(\symbSM_{c}, \symbpolicy_{c})$ is a security-preserving refinement of $(\symbSM_{a}, \symbpolicy_{a})$, represented as $(\symbSM_{c}, \symbpolicy_{c}) \refine (\symbSM_{a}, \symbpolicy_{a})$ , if and only if
\begin{equation*}
\begin{aligned}
   & LR(\symbSM_{a}, \symbpolicy_{a}) \implies LR(\symbSM_{c}, \symbpolicy_{c}) \\
   & SC(\symbSM_{a}, \symbpolicy_{a}) \implies SC(\symbSM_{c}, \symbpolicy_{c})
\end{aligned}
\end{equation*}
\end{definition}

\section{Unwinding-Preserving Simulation}
\label{sect:sim}

In this section, we introduce an unwinding-preserving simulation to verify the refinement relation. 
\subsection{Unwinding-Preserving Simulation}
Given the implementation $\symbSM_{c} = \langle \symbState, \symbAction_{c}, step_{c}, \symbstate_{0} \rangle$ with the security policy $\symbpolicy_{c} = \langle \symbDomain, \interf_{c}, dom_{c}, \sim_{c} \rangle$, and abstract specification $\symbSM_{a} = \langle \Sigma, \symbAction_{a}, step_{a}, \sigma_{0} \rangle$ with security policy $\symbpolicy_{a} = \langle \symbDomain, \interf_{a}, dom_{a}, \sim_{a} \rangle$, our defined unwinding-preserving simulation is in the form of: $ (\symbSM_{c}, \symbpolicy_{c}) \simuli_{\alpha, \zeta} (\symbSM_{a}, \symbpolicy_{a})$. The simulation takes two additional parameters: the state invariant relation $\alpha$ and the step mapping function $\zeta$. The state invariant relation $\alpha$ is the state relation between the implementation and the abstract specification. The step mapping function $\zeta$ is the step relation between the implementation and the abstract specification. If $\zeta(\symbaction_{c}) = \tau$, then $\symbaction_{c}$ is a silent step. Otherwise, if $\zeta(\symbaction_{c}) = \symbaction_{a}$, there is a corresponding step $\symbaction_{a}$ performed by the abstract specification. 

$$ \alpha \in \mathcal{P}(\symbState \times \Sigma) \quad \zeta: \symbAction_{c} \mapsto \symbAction_{a} \cup \{ \stutter \}$$

\begin{definition}[Unwinding-Preserving Simulation] 
\label{def:mu}
The simulation $ (\symbSM_{c}, \symbpolicy_{c}) \simuli_{\alpha, \zeta} (\symbSM_{a}, \symbpolicy_{a})$ is unwinding-preserving if and only if the following conditions hold:
\begin{enumerate}
    \item $(\lstate_{0}, \hstate_{0}) \in \alpha$.
    \item if $(\lstate, \hstate) \in \alpha$, $\lstate' \in step_{c}(\lstate, \symbaction_{c})$, and $\zeta(\symbaction_{c}) = \stutter$, then $(\lstate', \hstate) \in \alpha$.
    \item if $(\lstate, \hstate) \in \alpha$, $\lstate' \in step_{c}(\lstate, \symbaction_{c})$, and $\zeta(\symbaction_{c}) = \symbaction_{a}$, then $\exists \hstate'. \ \hstate' \in step_{a}(\hstate, \symbaction_{a})$ and $(\lstate', \hstate') \in \alpha$.
    \item if $\zeta(\symbaction_{c}) = \symbaction_{a}$, then $ dom_{c}(\symbaction_{c}) = dom_{a}(\symbaction_{a})$
    \item $\interf_{a} \ \subseteq \ \interf_{c}$.
    \item If $(\lstate_{1}, \hstate_{1}) \in \alpha$ and $(\lstate_{2}, \hstate_{2}) \in \alpha$, then $\lstate_{1} \ \stackrel{d}{\sim_{a}} \ \lstate_{2} \equiv \hstate_{1} \ \stackrel{d}{\sim_{c}} \ \hstate_{2}$ for all $\symbdomain$.
\end{enumerate}   
\end{definition}

Informally, $ (\symbSM_{c}, \symbpolicy_{c}) \simuli_{\alpha, \zeta} (\symbSM_{a}, \symbpolicy_{a})$ says the implementation $\symbSM_{c}$ with the security configuration $\symbpolicy_{c}$ simulates the abstract specification $\symbSM_{a}$ with the security configuration $\symbpolicy_{a}$. It requires the following hold for every execution of $(\symbSM_{c}, \symbpolicy_{c})$:

\begin{enumerate}
    \item The initial states are $\alpha$-related
    \item Each step $\symbaction_{c}$ of $\symbSM_{c}$ corresponds to either a silent step $\tau$ or a step $\symbaction_{a}$ in abstract specification, as determined by the step mapping function $\zeta$. The resulting states are $\alpha$-related too.
    \item The step mapping function preserves the domain of the action, for example, if a thread $t$ acts $\symbaction_{c}$, it must be mapped to a step $\symbaction_{a}$ that is also performed by thread $t$.
    \item The security policy in the implementation is stricter than in the abstract specification. This ensures that if a system is proven secure under a more relaxed policy, it remains secure under a stricter policy. If the security policies in the implementation and the abstract specification are identical, this condition holds trivially.
    \item State invariant relation preserves state indistinguishability.
\end{enumerate}

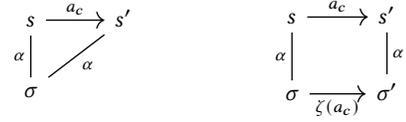
\begin{figure}
\begin{tabular}{cc}
    \begin{tikzcd}
        \symbstate \arrow[r, "a_{c}"] \arrow[d, dash, "\alpha"']
        & \symbstate' \arrow[ld, dash, "\alpha"] \\
        \sigma
    \end{tikzcd} 
     &  
    \begin{tikzcd}
        \symbstate \arrow[r, "a_{c}"] \arrow[d, dash, "\alpha"']
        & \symbstate' \arrow[d, dash, "\alpha"] \\
        \sigma \arrow[r, "\zeta(a_{c})"']
        & \sigma'
    \end{tikzcd}  \\
        (a) Silent Step $\left (\zeta (a_{c}) = \tau \right)$ & (b) Corresponding Step
\end{tabular}
\caption{Related Transitions}
\label{fig:step}
\end{figure}

Fig. \ref{fig:step} illustrates why simulation preserves unwinding conditions. If the action performed by the implementation is a silent step, then condition (2) requires that the resulting state after execution is also $\alpha$-related to the same abstract state. Since condition (6) requires that $\alpha$ preserves the state indistinguishability, we can conclude that the silent step will never break state indistinguishability. If the action $a_{c}$ performed by the implementation has a corresponding step $\zeta(a_{c})$, then condition (2) requires the resulting state after executing $a_{c}$ and $\zeta(a_{c})$ are still $\alpha$-related. Since condition (6) requires $\alpha$ to preserve the state indistinguishability, we know the corresponding step has the same effect (either both preserve state indistinguishability or both break it). Furthermore, as condition (5) suggests, since the domains performing the corresponding steps are the same, they will either both obey the security policy or violate it. 

For the example shown in Fig.~\ref{fig:example}, the state invariant ensures that the observable message queue remains consistent between the abstraction and the implementation. In the abstraction, this corresponds directly to the actual state of the message queue. In the implementation, however, it reflects the state of the queue as last observed— specifically, the state when the lock protecting the queue was last released. To capture this notion formally, we introduce a ghost field, $obq$, in the message queue data structure to represent the observable state of the queue in the implementation. That is, $q \texttt{->} obq$ denotes the observable state of message queue $q$. Its formal definition will be presented in Section~\ref{subsect:concrete}.

$$\alpha \defi \{(\symbstate, \sigma) \mid \ \forall q. \ s(q \texttt{->} obq) = \sigma(q \texttt{->} que) \}$$

The step mapping function is defined such that every step in the implementation is mapped to a silent step, except for the unlock action, which is mapped to the atomically enqueue action in the abstract specification. 

\[
\zeta(t, \symbprog_{c}) = 
\begin{cases}
(t, \text{send}), & \text{if } \symbprog_{c} = unlock \\
\stutter, & \text{otherwise}
\end{cases}
\]

The silent step does not break the state indistinguishability because, when the lock is held, other threads cannot observe the actual state of the message queue. However, when executing the unlock step, the message queue becomes observable to other threads, and we must ensure that the state invariant is re-established after executing the corresponding step.

We proved a soundness theorem showing that the unwinding-preserving simulation implies security-preserving refinement:
\begin{theorem}[Soundness of Simulation]
\label{thm:sound}
    If there exists $\alpha$ and $\zeta$ such that $(\symbSM_{c}, \symbpolicy_{c}) \simuli_{\alpha, \zeta} (\symbSM_{a}, \symbpolicy_{a})$, then $(\symbSM_{c}, \symbpolicy_{c}) \refine (\symbSM_{a}, \symbpolicy_{a})$
\end{theorem}

\subsection{Compositional Verification}
To simplify reasoning about all possible interleavings, we lift concurrent execution to a component-local model, which distinguishes execution within a specific parallel component from its concurrent environment. We use rely-guarantee-based simulation, similar to RGSim\cite{rgsim}, to enable compositional reasoning. Each parallel component $\symbcore$ is parametrized with the rely condition $\lRely_{\symbcore}, \hRely_{\symbcore}$,
which captures state transitions induced by the environment at the specification level and abstraction level, and a guarantee condition $\lGuar_{\symbcore}, \hGuar_{\symbcore}$, which describes state transitions performed by $\symbcore$ itself at the specification level and abstraction level. These conditions are defined at both the specification and abstraction levels as binary relations over states, enabling compositional reasoning:

$$ \lRely_{\symbcore}, \lGuar_{\symbcore} \in \mathcal{P}(\lstate \times \lstate) \quad \hRely_{\symbcore}, \hGuar_{\symbcore} \in \mathcal{P}(\hstate \times \hstate)$$

Each step in the component-local model is either a local step, whose effect is governed by the guarantee condition, or an environmental step, whose effect is captured by the rely condition. For example, in Fig. \ref{fig:example}, when the lock is held by the thread running on $CPU_\symbcore$, we can show that the environment (i.e., other CPUs) cannot modify the message queue, as enforced by the specification-level rely condition $\lRely_{\symbcore}$. Similarly, when the lock is held by a thread on another $CPU_{\symbcore'}$, the $CPU_\symbcore$ itself cannot access the message queue, as specified by the guarantee condition $\lGuar_{\symbcore}$.

If we can prove the simulation relation for each parallel component, and the rely-guarantee conditions are compatible across different parallel components (the guarantee condition of the parallel component implies the rely condition of others), then the simulation relation of the concurrent system can be proved. 

We decompose the simulation proof of the concurrent system (conditions (2) and (3) in definition \ref{def:mu}) into a set of lemmas focusing on individual parallel components. 
\begin{lemma}
If the parallel component $\symbcore$ makes a step $\symbaction_{c}$ from $\symbstate$ to $\symbstate'$, i.e., $\symbstate' \in step_{c}(\symbstate, \symbaction_{c})$, $(\symbstate, \sigma) \in \alpha$, $\zeta(\symbaction_{c}) = \stutter$, then $(\symbstate, \symbstate') \in \lGuar_{\symbcore}$ and $(\symbstate', \sigma) \in \alpha$.    
\end{lemma}
\begin{lemma}
If the parallel component $\symbcore$ makes a step $\symbaction_{c}$ from $\symbstate$ to $\symbstate'$, i.e., $\symbstate' \in step_{c}(\symbstate, \symbaction_{c})$, $(\symbstate, \sigma) \in \alpha$, $\zeta(\symbaction_{c}) = \symbaction_{c}$, then $\exists \sigma'.$ such that $\sigma' \in step_{a}(\sigma, \symbaction_{a})$, $(\symbstate, \symbstate') \in \lGuar_{\symbcore}$, $(\sigma, \sigma') \in \hGuar_{\symbcore}$ and $(\symbstate', \sigma') \in \alpha$.       
\end{lemma}
\begin{lemma}
If $(\symbstate, \sigma) \in \alpha$, $(\symbstate, \symbstate') \in \lRely_{\symbcore}$, and $(\sigma, \sigma') \in \hRely_{\symbcore}$, then $(\symbstate', \sigma') \in \alpha$.
\end{lemma}
\begin{lemma}
Rely and guarantee conditions between different parallel components are compatible, i.e., $\forall \symbcore, \symbcore'. \ \symbcore \neq \symbcore' \implies \lGuar_\symbcore \subseteq \lRely_{\symbcore'} \wedge \hGuar_{\symbcore} \subseteq \hRely_{\symbcore'}$
\end{lemma}

Lemmas 1 and 2 ensure that starting from $\alpha$-related states,  each step corresponds to either a silent step $\tau$ or a step $\symbaction_{a}$ in the abstract specification, as determined by the step mapping function $\zeta$. The resulting states are $\alpha$-related too. Furthermore, it requires the state transitions to be included in the guarantee conditions of both the implementation and the abstract specification.

Lemma 3 ensures that every environmental step also preserves the $\alpha$ relation. Lemma 4 guarantees that the rely and guarantee conditions between different parallel components are compatible, meaning that one component's assumptions about the environment are correctly upheld by other components.
\begin{theorem}[Soundness of Compositional Simulation]
    If Lemma 1 - 4 holds, then conditions (2) and (3) in Definition \ref{def:mu} hold.
\end{theorem}

For the sake of readability, we omit the proof system; the details of the proof system can be found in our Isabelle/HOL formalization.

\subsection{Example}
\label{subsect:concrete}
Below, we provide a simple example to illustrate the use of unwinding-preserving simulation and its parallel compositionality in verifying concurrent system refinement. The abstract specification of the send function atomically enqueues a message to the destination queue if it is not full.

The implementation introduces a lock $l$ to synchronize access to the shared message queues, enabling a more fine-grained version of the send function. Specifically, if a thread $t$ attempts to read a message from its message queue while another thread $t'$ holds the lock, $t$ cannot observe the message queue until $t'$ releases the lock.

\begin{figure}
    \centering
    \begin{multicols}{2}
    \begin{lstlisting}[style=CStyle]
struct msgq{
  size, max :: nat;
  que :: msg list}
struct kernel_state {
  cid :: tid;
  t_msg :: tid (* $\rightarrow$*) msgq}
void send(tid id, msg m){
  (*$\langle$*) if check(cid, id, policy)
    msgq q = t_msg(id);
    if (q->size < q->max)
      q->que[size] = m;
      q->size ++ (*$\rangle$*)}
    \end{lstlisting}   
    \end{multicols}
    \rotatebox{90}{$\refine$}
    \begin{multicols}{2}
    \begin{lstlisting}[style=CStyle]
struct msgq {
  size, max :: nat;
  que :: msg list;
  (*\textcolor{violet}{obq :: msg list;} *)
  mutex l}
struct kernel_state {
  cid :: tid;
  t_msg :: tid (* $\rightarrow$*) msgq
void send(tid id, msg m){
  if check(cid, id, policy)
    msgq q = t_msg(id);
    lock(q->l)
    if (q->size < q->max)
      q->que[size] = m;
      q->size ++;
    (* \textcolor{violet}{$\langle$ unlock(q->l); q->obq = q->que} $\rangle$ *)}
    \end{lstlisting}   
    \end{multicols}
    \caption{Unwinding-Preserving Simulation for Send Function}
    \label{fig:verify}
\end{figure}

To track each thread’s observations in the implementation, we introduce an auxiliary variable $obq$ which records the state of the message queue from the last time it was accessible. When the lock is released, the $obq$ accurately reflects the actual message queue. However, if the lock is held by a thread, $obq$ retains the state of the message queue from the last time the lock was released and may become inconsistent with the actual queue.

All auxiliary variables and auxiliary commands (highlighted in purple) are used solely for tracking observations and do not affect the actual execution of the implementation. We now define the observation in the abstract specification and implementation:

\begin{equation*}
\begin{aligned}&  \symbstate \stackrel{t}{\sim_{c}} \symbstate' \defi \symbstate(t\_msg(t) \texttt{->} obq) = \symbstate'(t\_msg(t) \texttt{->} obq) \\
&  \sigma \stackrel{t}{\sim_{a}} \sigma' \defi \sigma(t\_msg(t) \texttt{->} que) = \sigma'(t\_msg(t) \texttt{->} que) 
\end{aligned}
\end{equation*}

In the abstract specification, a thread can directly observe its message queue. However, in the implementation, a thread can only observe the state of the message queue from the last time it was accessible, as the message may be unavailable if another thread has acquired the lock.

To prove the simulation relation, we need to provide the state relation invariant $\alpha$ and the step mapping function $\zeta$. We begin by defining the state invariant $\alpha$, which ensures that the observable message queue remains consistent between the abstraction and the implementation. Specifically, when the lock is available, the observable message queue in the implementation accurately reflects the actual state of the message queue:

\begin{equation*}
\begin{aligned}
& \alpha \defi \{(\symbstate, \sigma) \mid \ \forall q. \ s(q \texttt{->} obq) = \sigma(q \texttt{->} que) \ \wedge \\
&  (s(q \texttt{->} l) = 0 \implies s(q \texttt{->} obq) = s(q\texttt{->} que)) \}   
\end{aligned}
\end{equation*}

It is straightforward to show that the state relation invariant $\alpha$ preserves the state indistinguishability relation:


The step mapping function is defined as  every step in the implementation is mapped to the silent step, except the unlock action, which is mapped to the atomically send action in the abstract specification:
\[
\zeta(t, \symbprog_{c}) = 
\begin{cases}
(t, \text{send}(id, m)), & \text{if } \symbprog_{c} = unlock \text{ of send}(id, m) \\
\stutter, & \text{otherwise}
\end{cases}
\]

It is straightforward to show that silent steps do not break the state invariant relation, as they do not modify the value of $obq$. The $unlock$ action atomically releases the lock and updates $obq$ with the actual state of the message queue, ensuring that the state invariant relation is preserved after the send action is performed in the abstract specification.  Furthermore, it is easy to verify that both non-silent steps belong to the same domain, specifically, the thread that invokes the send function. 

To apply compositionally verification, we apply rely-guarantee simulation. The abstraction-level threads can execute the send function in arbitrary environments with arbitrary guarantees: $\hRely_{\symbcore} = \hGuar_{\symbcore} \defi True$. The implementation uses the lock to protect every access to message queues. Therefore, the specification-level relies and guarantees are not arbitrary:

\begin{equation*}
\begin{aligned}
& \lRely_{\symbcore} \defi \{(\symbstate, \symbstate') \mid \forall q. \ \symbstate(q \texttt{->} l) = \symbcore \implies  \\
& \symbstate(q \texttt{->} que) = \symbstate'(q \texttt{->} que) \wedge \symbstate(q \texttt{->} obq) = \symbstate'(q \texttt{->} obq) \wedge \\
& \symbstate(q \texttt{->} size) = \symbstate'(q \texttt{->} size) \} \\
& \lGuar_{\symbcore} \defi \{ (\symbstate, \symbstate') \mid \ \symbstate = \symbstate' \ \lor \ (\symbstate(q \texttt{->} l) = 0 \wedge \symbstate'((q \texttt{->} l) \leadsto \symbcore)) \\ 
& \lor (\symbstate (q \texttt{->} l) = \symbcore \wedge \symbstate' = \symbstate ((q \texttt{->} que) \leadsto -, (q \texttt{->} size) \leadsto -)) \\
& \lor (\symbstate (q \texttt{->} l) = \symbcore \wedge \symbstate' = \symbstate ((q \texttt{->} l) \leadsto 0, (q \texttt{->} obq) \leadsto -)) \}
\end{aligned}
\end{equation*}

Every CPU guarantees that it updates the message queue only when the lock is acquired. Its environment cannot update the message queue if the CPU currently holds the lock. It is easy to prove rely and guarantee conditions between different CPUs are compatible:


As shown in Theorem \ref{thm:sound}, the simulation is unwinding-preserving, so if the send function is proved secure in the abstract specification, then so is the send function in the implementation.

\section{Verification of IPC Mechanism}
\label{sect:case}

In this section, we provide our formalization of the IPC mechanism for the ARINC 653 multicore standard. We identified and rectified two IFS violations during the verification process. 
\subsection{Overview of ARINC 653 Multicore} 

\begin{figure}[t]
    \centering
    \includegraphics[width=3in]{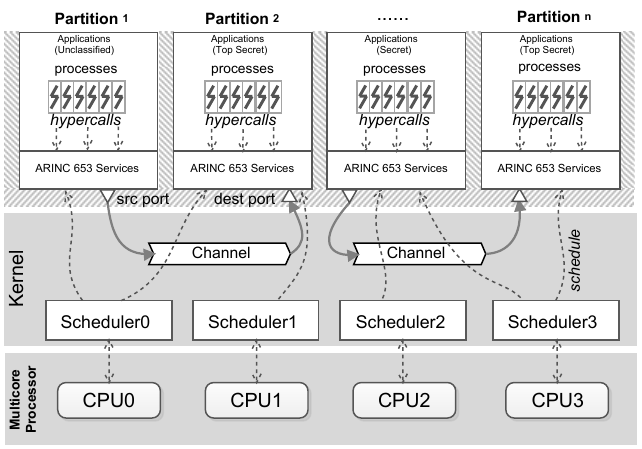}
    \caption{Architecture of Multicore Separation Kernels}
    \label{fig:arch}
\end{figure}

The ARINC 653 standard - Part 1 in Version 5 \cite{ARINC653p1_4} released in 2019 specifies the baseline operating environment for application software used within Integrated Modular Architecture on a multicore platform. It defines the system functionality and requirements of system services for separation kernels. The ARINC 653 standard was originally written in APEX pseudocode and natural language. In our case study, we manually translated it into our modeling language (see the appendix). 

IFS in separation kernels ensures no information flows between partitions except through defined communication channels. The security policy used by separation kernels is the Inter-Partition Flow Policy (IPFP) \cite{IPFP}, which is intransitive. It is expressed abstractly in a partition flow matrix $\mathbf{partition\_flow}: partition \times partition \rightarrow mode$, whose entries indicate the mode of the flow. For instance, $\mathbf{partition\_flow}(\symbpart_1,\symbpart_2) = QUEUING$ means that a partition $\symbpart_1$ is allowed to send information to a partition $\symbpart_2$ via a message queue. The information flow relation between partitions can be derived from the partition flow matrix straightforwardly.

\subsection{System Specification}

Fig. \ref{fig:arch} shows the outline of the ARINC 653 multicore, which is the parallel composition of multiple single-core systems. Each single-core system in ARINC 653 contains one scheduler and multiple partitions. Partitions function similarly to processes in Windows or Unix systems, capable of continuously invoking services from the underlying operating systems via hypercalls. Our case focuses on partitions, partition scheduling, and inter-partition communication (IPC) through queuing channels. IPC is conducted via messages on channels configured among partitions. Partitions have access to channels via ports, which are the endpoints of channels.

We instantiate the security model for the ARINC 653 multicore system with the following concrete parameters:

\paragraph{System State} A significant characteristic of ARINC 653 is that the basic components are statically configured at build time. The configuration are defi as follows ($\to$ means a total mapping):

\begin{itemize}
    \item $c2s :: CPU \to Sched$ --- A mapping from CPUs to schedulers and is bijective.
    \item $p2s :: Part \to Sched$ --- The deployment of partitions to schedulers.
    \item $p2p :: Port \to Part$ --- The deployment of communication ports to partitions
    \item $chsrc :: QChannel \to Port$, $chdest :: QChannel \to Port$ --- The source port and destination port of a queuing channel.
    \item $chmax :: QChannel \to nat$ --- The capacity of a queuing channel.
\end{itemize}

The dynamic component contains components as follows:

\begin{itemize}
    \item $cur :: Sched \rightharpoonup Part$ --- The current partition under execution ($\rightharpoonup$ means the partial mapping).
    \item $partst :: Part \rightarrow PartMode$ --- The running state of a partition which is either IDLE (can not be scheduled), READY (ready to be scheduled), or RUN (currently scheduled).
    \item $qbuf :: QChannel \to Message \text{ list}$ --- the message list in the queuing channel.
    \item $qbufsize :: QChannel \to nat$ --- The amounts of messages in the queuing channel.
\end{itemize}

The implementation further applies $qlock$ to synchronize the accesses of the shared message queues and has a ghost variable $obuf$ to track the observation of each thread, which we already explained in Section \ref{sect:sim}:

\begin{itemize}
    \item $qlock :: QChannel \rightharpoonup CPU$ --- The lock of a queuing channel
    \item $obuf :: QChannel \to Message \text{ list}$ --- The observation of a queuing channel.
\end{itemize}

\paragraph{System Call} 

In this case study, we model partitions, partition scheduling, and inter-partition communication using queuing channels. Our analysis considers four handler programs, each represented in the form: 
$$\text{ProgName } [\text{Parameters}] \ @ \text{ CPU}$$ 

where ProgName denotes the handler program’s name, Parameters specifies its input arguments, and CPU indicates the processing unit on which the handler is executed.
\begin{itemize}
    \item $Core\_Init \ [ \ ] \ @ \ \symbcore$ --- Initializes the running state of partitions deployed on CPU \ $\symbcore$.
    \item $Schedule \ [Part] \ @ \ \symbcore$ --- Schedules the partition $Part$ on the CPU $\symbcore$. It requires the parameter partition $Part$ to be deployed on the CPU $\symbcore$ and that $Part$ is not in the $IDLE$ state. 
    \item $Recv\_QMsg \ [p] \ @ \ \symbcore$ --- Receives messages from port $p$ on CPU $\symbcore$. It requires the parameter port $p$ to be configured as the destination port and is deployed on the currently running partition on CPU $\symbcore$.
    \item $Send\_QMsg \ [p, m] \ @ \ \symbcore$ --- Sends message $m$ through port $p$ on CPU $\symbcore$. It requires the parameter port $p$ to be configured as a source port and is deployed on the currently running partition on CPU $\symbcore$. 
\end{itemize}

The entire multicore system is modeled as a parallel composition of multiple single-core systems, each containing a scheduler and several partitions, and parameterized by a CPU identifier $\symbcore$. The system begins with the event $Core\_Init$, which initializes the kernel state on each CPU. Following this, the kernel instance responds to system calls, including schedule and IPC functions invoked continuously throughout execution.

\paragraph{Security Policy}
There are two kinds of security domains for ARINC 653 multicore, represented as:
\begin{equation*}
\begin{aligned}
\symbDomain \defi & \ \{\symbpart_{11}, \cdots, \symbpart_{1j}, \cdots \symbpart_{\symbcore1}, \cdots, \symbpart_{\symbcore n} \} \\
& \cup \{sched_{1} \cdots sched_{\symbcore}\}  
\end{aligned}
\end{equation*}

$\symbpart_{\symbcore j}$ is a partition domain that is deployed on CPU $\symbcore$, responsible for performing IPC functions. $sched_{\symbcore} $ is the scheduler domain deployed on CPU $\symbcore$, responsible for initialization and scheduling.

The security policy in both the abstract specification and the implementation remains the same. It is defined based on the channel configuration. Information can always flow to itself. The scheduler can schedule all partitions deployed on the same CPU, so information can flow from the scheduler to these partitions. To ensure that the scheduler does not leak information via its scheduling decisions, no information can flow to a scheduler except from itself. If there is a channel from a partition $\symbpart_{ij}$ to a partition $\symbpart_{i'j'}$, i.e., a channel's source port is deployed on $\symbpart_{ij}$, and the destination port is deployed on $\symbpart_{i'j'}$, then $\symbpart_{ij} \interf \symbpart_{i'j'}$. The security policy is defined as follows:

\begin{enumerate}
    \item $\forall \symbdomain \in \symbDomain. \ \symbdomain \interf \symbdomain$
    \item $sched_{\symbcore} \interf \symbpart_{ki}$
    \item $\symbdomain \interf sched_{\symbcore} \implies \symbdomain  = sched_{\symbcore}$
    \item $\symbpart_{ij} \interf \symbpart_{i'j'}$ if and only if there is a channel from $\symbpart_{ij} $ to $\symbpart_{i'j'}$
\end{enumerate}

\paragraph{State Indistinguishability Relation}
We now define the state indistinguishability relation used in our verification:

\begin{itemize}
    \item Scheduler --- The scheduler $sched_{\symbcore}$  can observe the currently running partition on that CPU. All states with the same running partition on that CPU can not be distinguished by the scheduler.
    \item Partition --- A partition can observe its own running state and the messages it receives. Specifically, if a channel $q$'s destination port is deployed on the partition $\symbpart$, then partition $\symbpart$ can observe messages in that channel. The key difference is that, in the abstract specification, the partition can observe the actual message values, whereas in the implementation, it can only observe the channel's state as it was the last time it was accessible.
\end{itemize}

\paragraph{Security Proof}
The key to simulation proof is to provide a state relation invariant $\alpha$, and step mapping function $\zeta$, and prove they both preserve security.

In our case study, the state relation invariant $\alpha$ ensures that the current running partition and the state of the partition remain the same in the abstract specification and implementation. For the message queue, it ensures the actual value of the message queue in the abstract specification is consistent with the state of the message queue from the last time it was accessible in the implementation.

For kernel initialization and schedule functions, the step mapping function is defined as an identity map from the implementation to the abstract specification. For IPC functions, the step mapping function is defined as every step in the implementation is mapped to the silent step, except the unlock action, which is mapped to the atomically enqueue action in the abstract specification for the send function, and is mapped to the atomically dequeue action in the abstract specification for the recv function.
\[
\zeta(t, \symbprog_{c}) = 
\begin{cases}
(t, \text{send}(id, m)), & \text{if } \symbprog_{c} = unlock \text{ of send(id, m)} \\
(t, \text{recv}()), & \text{if } \symbprog_{c} = unlock \text{ of recv()} \\
\stutter, & \text{otherwise}
\end{cases}
\]

We proved that silent steps do not break the state invariant relation. The step mapping for kernel initialization and scheduling functions is the identity map that preserves the state invariant trivially. The $unlock$ action atomically releases the lock and atomically releases the lock and updates $obuf$ with the actual state of the message queue, ensuring that the state invariant relation is preserved after the enqueue or dequeue action is performed in the abstract specification. Furthermore, it is easy to verify that both non-silent steps belong to the same domain—specifically, the same scheduler invokes the kernel initialization and scheduling functions, and the same thread invokes the IPC function. 

Then we prove the security-preserving refinement between the abstraction and the implementation. We prove the unwinding conditions hold for the abstract specification. By Theorem \ref{thm:sound}, we show the implementation also satisfies the unwinding conditions.

\section{Verification of Declassification}
\label{sect:case1}

Practical programs with sensitive data often need to reveal part of it intentionally. The declassification policy is used to specify such systems that ensure only intentionally released information can be learned by others.  Our second case study examines a sealed-bid auction, adapted from \cite{verdeca}. In this type of auction, all bids remain confidential until the auction concludes, preventing bidders from adjusting their bids in response to others. 

\subsection{Overview of Sealed-bid Auction}
The auction process is managed by a bid server, which is responsible for starting and closing the auction. Each client bid is processed by a dedicated bid-handling thread. A client can submit a bid once the auction has started; the system then compares the new bid with the current maximum and updates it if the new bid is higher. Bids are represented as pairs $(id, qt)$ where $id$ is the identity of the client who submitted the bid and $qt$ is the amount (or quote) of the bid. To ensure fairness, bid-processing threads execute concurrently, preventing any bidder from denying service to others. Once the auction concludes, the result publisher announces the outcome. The security policy guarantees that bidders cannot access information about competing bids until the results are published. Moreover, the only information revealed to bidders throughout the auction is the identity of the highest bidder.

\subsection{System Specification}

We instantiate the security model for the seal-bid auction system with the following concrete parameters:

\paragraph{System State}  

The state of the auction system contains the following components:

\begin{itemize}
    \item $status :: Auc\_State$ --- The current state of the auction. The auction can be in one of three states: CLOSED (indicating that the auction has ended), READY (indicating that the auction is prepared to receive bids), or RUNNING (indicating that bidding is actively in progress).
    \item $reserve :: nat$ --- The auction's reserve price.
    \item $max\_bid :: uid \times nat$ --- The current highest bid along with the corresponding bidder.
    \item $log :: uid \times nat \text{ list}$ --- Maintains a history of all registered bids, serving as an audit trail in case of disputes.
    \item $res :: Auc\_Res$ --- The auction’s result, which is either the highest bid or unsuccessful if no bid exceeds the reserve price.
\end{itemize}

In the implementation, we further have $lock$ to synchronize the accesses of the shared resources. It further has ghost variables $olog$ and $obid$ to track the observation:

\begin{itemize}
    \item $lock:: CPU \text{ option}$ --- The lock of the shared variables. 
    \item $obid :: uid \times nat$ --- The observation of the highest bid.
    \item $oblog :: uid \times nat \text{ list}$ --- The observation of the bids history.
\end{itemize}

\paragraph{System Call} 
We consider four handler programs in our case study:

\begin{itemize}
    \item $Start\_Auction \ [ qt ] \ @ \ Auction$ --- Initiates the auction and sets the reserve price to $qt$.
    \item $Close\_Auction \ [ \ ] \ @ \ Auction$ --- Closes the auction.
    \item $Publish\_Result \ [ \ ] \ @ \ Auction$ --- Publishes auction results. 
    \item $Register\_Bid \ [id, qt] \ @ \ (User \ id)$ --- Registers a bid of amount $qt$ from a user identified by $id$. 
\end{itemize} 

The entire multicore system is modeled as a parallel composition of multiple user threads and the auction thread. A user's thread can continuously register bids while the auction is running. The auction thread can start and close the auction, as well as publish the auction results once it has ended.
 
\paragraph{Security Policy}
There are three kinds of security domains for our auction example, represented as:
\begin{equation*}
\begin{aligned}
\symbDomain \defi & \ \{User_{1}, \cdots, User_{n} \} \cup \{ Server \} \cup \{ Publisher \} 
\end{aligned}
\end{equation*}

$User_{i}$ is a user domain responsible for performing register functions. $Server$ is the server domain deployed on the auction holder's computer, responsible for starting and closing the auction. $Publisher$ is the publisher domain deployed on the auction holder's computer, responsible for publishing the result of the auction.

The security policy remains the same in both the abstract specification and the implementation. Information can always flow to itself. Information can flow from the user domain to the server domain when a user registers a new bid. However, information cannot flow directly from the server to the users. Any information transfer from the server domain to the user domain must be mediated by the publisher domain. This ensures that all information released from the server to the users is properly downgraded by the publisher. The security policy is defined as follows:

\begin{enumerate}
    \item $\forall \symbdomain \in \symbDomain. \ \symbdomain \interf \symbdomain$
    \item $User_{i} \interf Server$
    \item $Server \interf Publisher$
    \item $Publisher \interf User_{i}$
\end{enumerate}

\paragraph{State Indistinguishability Relation}

We now define the state indistinguishability relation used in the auction system:
    \textbf{1) User and Publisher} --- The user and publisher can only observe the result of the auction, ensuring users cannot access information about competing bids.
    \textbf{2) Auction Server} --- The server can observe all information about the auction. The key difference is that, in the abstract specification, the server can observe the actual value of the log and the maximum bid, whereas in the implementation, it can only observe the log and the maximum bid as they were the last time they were accessible.

\subsection{Security Proof}

In this case study, the state relation invariant $\alpha$ ensures the auction status, reserve price, and result remain the same in the abstract specification and implementation. For the bid log and max bid, it ensures the actual value of the log and max bid in the abstract specification is consistent with the state of the log and max bid from the last time it was accessible in the implementation. Further, we ensure the result will be the max bid only when the auction is closed, and the max bid is greater than the reserve price.

The state invariant, together with the security policy, ensures that no bid information can be declassified until after the auction is closed. Moreover, the only bid information that can be released is the maximum bid, but only if it exceeds the reserve price.


\section{Evaluation}
\label{sect:exp}
This section reports our experience applying our framework to verify the two case studies.
\subsection{Covert Channel Discussion}

We identified that two covert channels previously reported in the ARINC 653 single-core standard \cite{arinc653_ifs} also exist in the ARINC 653 multicore standard.

\begin{enumerate}
    \item \textbf{Queuing mode channel}: If there is a queuing channel from a partition $a$ to a partition $b$ and no channels from $b$ to $a$, then $a \interf b$ and $b \ninterf a$. In the original ARINC 653 standard, invoking $Send\_QMsg$ returns $NOT\_AVAILABLE$ or $TIMED\_OUT$ when the queue is full and $NO\_ERROR$ when the queue is not. However, the full status of the queue can be changed by $b$ through invoking $Recv\_QMsg$ and thus violating the local respect condition.
    \item \textbf{Port identifier channel}: In $Send\_QMsg$ and $Recv\_QMsg$ functions, there is no validation to ensure that the accessed port belongs to the calling partition. In this case, programs in a partition can guess the port identifiers of other partitions and then manipulate the ports. Therefore, the port identifier in ARINC 653 is a covert channel. As a result, the port identifier itself can serve as a covert channel. 
\end{enumerate}

To eliminate these covert channels, we introduced two modifications: (1) allowing message loss in the queuing mode channel to prevent other partitions from observing the queue’s full status, and (2) enforcing a check to ensure that the accessed port belongs to the current partition, thereby preventing the port identifier channel. We proved that the revised version adheres to the IFS properties.

\subsection{Proof Effort}
From our experience, proving unwinding-preserving simulation is not significantly different from standard simulation techniques. The key difference lies in the need to establish that the state invariant preserves state indistinguishability and to explicitly define a step mapping function. However, as our example shows, for some lock-based systems and handler programs with a clear linearization point, the state invariant clearly preserves state indistinguishability, and the step mapping function can be easily identified.

Table~\ref{table:effort} summarizes our proof effort for two case studies. The simulation relation includes both the state invariant and the step mapping function. The security check ensures that the state invariant preserves the state indistinguishability relation and the step mapping function preserves the domain of each action. We spent four weeks verifying the IPC mechanism of the ARINC 653 multicore standard and three weeks verifying the sealed-bid auction example. Our findings suggest that extending existing proofs based on standard simulation techniques to support security preservation does not impose a significant additional proof burden.

\begin{table}[]
\caption{Lines of code for the two verified systems}
\label{table:effort}
\begin{tabular}{lll}
componet & ARINC 653 & Auction \\
\textbf{system spec:} &  &  \\
implementation & 260 & 202 \\
abstraction & 140 & 137 \\
\hline
\textbf{proof input:} &  & \\
simulation relation & 57 & 53 \\
security check & 48 & 46 \\
simulation proof & 848 & 620\\
\hline
\end{tabular}
\end{table}

\subsection{Discussion on Limitations}

Our formalization of security-preserving refinement imposes several restrictions that simplify reasoning. First, Definition \ref{def:mu} does not allow stuttering at the abstraction level—i.e., steps in the implementation that correspond to no state change in the abstract model. Such stuttering behavior may be present in compiler verification, where low-level steps may correspond to a stutter step in high-level source programs, and synchronous models of noninterference \cite{sync,sync1}, where the absence of progress can leak information. This work focuses on covert storage channels rather than timing or progress-based leakage where abstractions involving stuttering can typically be replaced with equivalent, simpler abstractions that do not include stutter steps.

Second, our approach does not support domain-visible nondeterminism in the abstraction yet. This restriction is necessary to ensure that security properties are preserved from the abstract specification to the implementation. In general, a deterministic but insecure implementation may refine a secure, nondeterministic abstraction. For example, consider a secret boolean $x$ and a program $P$ that randomly prints true or false, independently of $x$. While $P$ is secure, it can be refined by an insecure program $Q$ that directly prints $x$. To prevent this, our framework adopts the same strategy used in seL4 \cite{sel4_noninterference,sel4_noninterference2} and CertiKOS \cite{mcertikos,serval}, which rules out $P$ as a valid secure program. The SC condition requires that if two outcomes $s_1, s_2 \in step(s, a)$, then they must be indistinguishable: $s_{1} \stackrel{d}{\sim} s_{2}$. Consequently, $P$ is not provably secure in our framework as two executions may yield distinguishable outcomes and thus violate the SC condition. While this restriction limits expressible nondeterminism, the successful verifications of seL4 \cite{sel4_noninterference,sel4_noninterference2}, CertiKOS \cite{mcertikos,serval}, NiKOS \cite{nickel}, and ARINC 653 \cite{arinc653_ifs,zhao_refinement} demonstrate that it does not significantly hinder practical usability. 
 
\section{Related Work}
\label{sect:relate}

\paragraph{Security-Preserving Refinement}
As explained in Sections \ref{sect:case}, standard refinement techniques \cite{rgsim,csim,Igloo,ccal,lili1,lili2,termpreserve1,termpreserve2,trillium,simuliris} for concurrent systems may fail to preserve security. Security-preserving refinement has been widely adopted in systems such as seL4 \cite{sel4_noninterference,sel4_noninterference2}, CertiKOS \cite{mcertikos,serval}, NiKOS \cite{nickel}, and ARINC 653 \cite{arinc653_ifs,zhao_refinement}. However, these works focus only on sequential systems and do not generalize to concurrent systems, where refinement can obscure unintentional information leakage to concurrent observers.

For security refinement in concurrent systems, \cite{Verismo,li2022design} transform security properties into safety properties and prove that refinement preserves those safety properties. However, \cite{Verismo} supports only transitive policies and cannot model downgrading processes involving intentional data release. Meanwhile, \cite{li2022design} encodes downgrading through a set of rules for intentional data release, but it remains unclear what kind of security guarantees are provided. 

Baumann et al.~\cite{baumann} propose an ignorance-preserving refinement framework for synchronous multi-agent systems and identify six desirable properties for security-preserving refinement. Our approach satisfies properties (1) support for different data representations, (3) partial support for declassification (through intransitive policy), (4) compositionality (including vertical and parallel composition, though not sequential), and (6) practical verification methods. The key difference lies in the target models and underlying security assumptions: while ~\cite{baumann} focuses on synchronous multi-agent systems without interleaving and uses a knowledge-based security model, our work targets interleaving-based concurrent systems and adopts intransitive noninterference as the security criterion.

The most relevant work is SeKVM \cite{sekvm,sekvm1,kvm_secure}, which introduces transparent trace refinement to successfully propagate unwinding conditions in concurrent systems. However, SeKVM is limited to isolation policies. To support intentional data release, it modifies the semantic model by replacing declassified sensitive data with non-sensitive data, making the security guarantees for the actual system unclear. In contrast, our approach provides a refinement technique that preserves unwinding conditions for intransitive policies. Consequently, we encode declassification policies directly as intransitive policies, avoiding semantic modifications while ensuring that real-world systems satisfy key information-flow security properties, such as intransitive noninterference \cite{rushby81} and TA-security \cite{otherifs}.

\paragraph{Language-Based Security} A large body of work \cite{sabelfeld_language,murray_compositional,covern,seccsl,seloc,coughlin_relyguarantee,commcsl} has explored verification of information-flow security in concurrent programs. These approaches prevent implicit information flow and side-channel leakage (e.g., timing and termination channels) by enforcing security through specific programming language constructs. However, they focus only on transitive policies and cannot model downgrading processes involving intentional data release. Recent works \cite{veronica,verdeca} adapt a knowledge-based security model to support intentional data release. However, these approaches do not address refinement, which may limit their scalability to large codebases. In contrast, our framework is language-agnostic and supports layer-by-layer refinement without requiring system reimplementation. Moreover, our refinement technique can also be applied to language-based security by ensuring that every declassification command in the implementation is mapped to a corresponding step in the abstraction, while all other steps remain silent, preserving state indistinguishability.

However, language-based approaches offer advantages in verifying timing channels, and some \cite{seccsl,verdeca} support automated reasoning. Our approach does not yet handle timing channels and relies heavily on interactive theorem provers, which demand significant manual effort for verifying complex systems.  
\section{Conclusion and Future Work}
In this paper, we propose a framework for verifying the information-flow security of concurrent systems under general security policies. We formalize IFS properties as unwinding conditions and introduce a compositional approach to prove that these conditions are preserved between implementation and abstraction. To demonstrate the effectiveness of our methodology, we apply it to verify two case-study systems against different security policies. Our proofs are fully mechanized in Isabelle/HOL, during which we identified and corrected an IFS violation. We plan to apply our framework to other multicore OS with IPC mechanisms as CertiKOS with IPC \cite{certikos} already provides a refinement proof but does not address security. 

\begin{acks}
We thank the anonymous reviewers for their insightful comments and constructive feedback. This work was supported by the Key R\&D Program of Zhejiang (No. 2025C01083), the Fundamental Research Funds for the Central Universities (No. 2025ZFJH02), and the Singapore Ministry of Education (No. MOE-T1-1/2022-43).
\end{acks}

\bibliographystyle{ACM-Reference-Format}
\balance
\bibliography{IFS}

\appendix 

\section{The Language}

\begin{figure}[thb]
\footnotesize
\begin{equation*}
\begin{aligned}
\text{(CoreId) } & \symbcore \in \mathtt{N} \qquad \qquad \qquad \quad\text {(State) } \symbstate \in \text {Var} \mapsto \text{Val} \\
\text{(Prog) } & \symbprog ::= \cmdnone \ | \ \cmdbasic{f} \ | \ \cmdseq{\symbprog_{1}}{\symbprog_{2}} \ | \ \cmdcond{b}{\symbprog_{1}}{\symbprog_{2}} \ |\\
& \cmdwhile{b}{\symbprog} \ | \ \cmdawait{b}{\symbprog} \ | \ \cmdnondt{r} \\
\text{(EvtLab) } &  \symblabel \in \text{Label} \qquad \qquad \quad \text{(Evtgrd) } g \in \mathcal{P}(\symbState)\\ 
\text{(Evt) } & \symbevt ::= (\symblabel, \symbguard, \symbprog) \qquad \text{(EvtPool) } \Pi \in \text{CoreId} \mapsto \mathcal{P}(\text{Evt}) \\
\text{(EvtDom) } & \symbdomf \in \text{Evt} \mapsto \symbDomain \quad \text{(EvtCtx) } \symbevtctx \in \text{CoreId} \mapsto \text{Evt} \\
\text{(Sys) } & \symbSys :: = \sysdefault \\
\text{(Config) } & \symbconf ::=  (\symbSys, \symbstate, \symbevtctx) \qquad \text{(Action) } \symbaction ::= (\symbdomain, \symbevt) \ | \ (\symbdomain, \symbevt, \symbprog)
\end{aligned}
\end{equation*}
\caption{The language syntax and state model
}
\label{fig:syntax}
\end{figure}   

To model the concurrent system, we employ an event-based programming language defined in Fig. \ref{fig:syntax} inspired by PiCore \cite{picore}. A concurrent system $\symbSys$ consists of an event pool $\Pi$ and multiple parallel subsystems executing handler programs. The event pool $\Pi$ maps each subsystem to a set of events, defining the event handlers the subsystem can execute (e.g., system calls or scheduling functions in a multicore OS). Each event is represented as a tuple $(\symblabel, \symbguard, \symbprog)$, where $\symblabel$ is the identifier encoding the name and parameters of the event, $\symbguard$ is the triggering condition that specifies parameter requirements, and $\symbprog$ is the handler program to be executed. Given an event $\symbevt$, we refer to its components as $\symbevt_{\symblabel}$, $\symbevt_{\symbguard}$ and $\symbevt_{\symbprog}$, respectively. The event domain function $\symbdomf: \text{Evt} \times \symbState \rightarrow \symbDomain$ maps an event handler and state to its corresponding domain. For instance, the domain of a system call corresponds to the thread that invoked it, while the domain of a scheduling function is the scheduler. A state-dependent $\symbdomf$ function is necessary for reasoning about many systems in which the domain of a system call depends on the currently running thread or process \cite{nickel, sel4_noninterference,arinc653_ifs}.

The operational semantics are defined through labeled configuration transitions: $ \symbconf \tran{\symbaction} \symbconf'$. The configuration $\symbconf$ is a tuple $(\symbSys, \symbstate, \symbevtctx)$ is a configuration,
where $\symbSys$ denotes the concurrent system, $\symbstate$ represents the system state, and $\symbevtctx$ specifies the event context, indicating which event is currently executed in the subsystem. Given a configuration $\symbconf$, we refer to its components as $\symbconf_{\symbSys}$, $\symbconf_{\symbstate}$ and $\symbconf_{\symbevtctx}$, respectively.
$\symbaction$ is the action stimulating the transition. In our context, an action can either be the invocation of a new event $\symbevt$ performed by domain $\symbdomain$, represented as $(\symbdomain, \symbevt)$, or an atomic step execution $P$ of the unfinished handler program performed by domain $\symbdomain$, represented as $(\symbdomain, \symbevt, \symbprog)$.

\section{Formalizing IFS for Concurrent Systems}

Intuitively, noninterference ensures that actions performed by a domain can only influence other domains to which it is permitted to transfer information (dictated by the security policy). 
Violation of noninterference thus means actions from a higher security domain cause observation changes at a lower security domain, enabling unauthorized information flow.

A purged trace is a sequence of system actions from which all actions related to high-security domains have been removed. Noninterference requires that the observations in the original trace and the purged trace remain identical for the domain, ensuring no unauthorized changes in the observation will be triggered. 
To construct a purged trace, we first define  $sources(\symbactions,\symbdomain)$ to identify the set of domains that are allowed to transfer information to domain $\symbdomain$ over the action trace $\symbactions$. 

\begin{equation*}
\left\{
\begin{aligned}
& sources ([ \ ], \symbdomain) = \{\symbdomain\} \\
& sources (\symbaction \# \symbactions, \symbdomain) = 
\begin{aligned}
& sources (\symbactions, \symbdomain) \cup \{w.\ w = dom(\symbaction) \ \wedge \\
& (\exists v. \ (w \interf v) \wedge v \in sources(\symbactions, \symbdomain))\}
\end{aligned}
\end{aligned}
\right.   
\end{equation*}

Then we use the function $ipurge(\symbactions,\symbdomain)$ to obtain the purged trace that retains the actions whose domain is identified by the function $sources$. 

\begin{equation*}
\left\{
\begin{aligned}
& ipurge ([ \ ], \symbdomain) = [ \ ] \\
& ipurge (\symbaction \# \symbactions, \symbdomain) = 
\begin{aligned}
& \mathbf{if} \ \symbdomain_\symbaction \in sources(\symbaction \# \symbactions, \symbdomain) \ \mathbf{then} \\
& \quad \symbaction \# ipurge (as, \symbdomain) \\
& \mathbf{else} \ \ ipurge (\symbactions, \symbdomain) 
\end{aligned}
\end{aligned}
\right.
\end{equation*}

Given a system $\symbSM =\langle \symbConf, \symbAction, step, \symbconf_0 \rangle$ and its associated information flow configuration $\symbpolicy = \langle \symbDomain, \interf, dom, \sim \rangle$, intransitive noninterference is formally defined as:
\begin{equation*}
\begin{aligned}
& \text{NI}(\symbSM, \symbpolicy) \defi \forall \symbactions, \symbdomain. \ \ssequidom{run(\symbconf_0, \symbactions)}{\symbdomain}{run(\symbconf_{0}, ipurge(\symbactions,\symbdomain))}
\end{aligned}    
\end{equation*}
Where
$$\ssequidom{\symbConf_{1}}{\symbdomain}{\symbConf_{2}} \defi \forall \symbconf1 \in \symbConf_{1}, \symbconf2 \in \symbConf_{2}. \ \symbconf1_{s} \dsim{\symbdomain} \symbconf2_{s}$$
The intuition of noninterference is that a domain $d$ cannot distinguish the final states produced by executing a sequence of actions $as$ and executing its purged sequence.

\end{document}